\documentclass{article}

\usepackage{arxiv}

\usepackage[utf8]{inputenc} 
\usepackage[T1]{fontenc}    
\usepackage{hyperref}       
\usepackage{url}            
\usepackage{booktabs}       
\usepackage{amsfonts}       
\usepackage{nicefrac}       
\usepackage{microtype}      
\usepackage{graphicx}
\usepackage[numbers]{natbib}
\usepackage{doi}

\usepackage{graphicx}
\usepackage{dcolumn}
\usepackage{bm}
\usepackage{hyperref}
\usepackage[mathlines]{lineno}
\usepackage{tensor}
\usepackage{newpxtext,newpxmath}

\usepackage{amsmath,amssymb}
\usepackage{quiver}

\newcommand{\beqa}{\begin{eqnarray}}
\newcommand{\eeqa}{\end{eqnarray}}
\newcommand{\p}{\partial}
\newcommand{\nn}{\nonumber}

\DeclareMathOperator{\diag}{diag}

\title{Anisotropic Conformal Dark Gravity on the Lorentz Tangent Bundle Spacetime}

\author{ {Christos Savvopoulos} \\
	Department of Mathematics\\
	Rheinische Friedrich-Wilhelms-Universität Bonn\\
	Endenicher Allee 60, 53115, Bonn, Germany \\
	\texttt{chsavopoulos@uni-bonn.de} \\
	\And
	\href{https://orcid.org/0000-0002-1187-8017}{\includegraphics[scale=0.06]{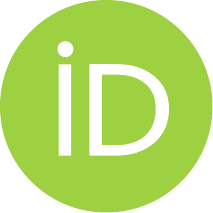}\hspace{1mm}Panayiotis Stavrinos} \\
	Department of Mathematics\\
	National and Kapodistrian University of 
Athens\\
	Panepistimiopolis, 15784, Athens, Greece \\
	\texttt{pstavrin@math.uoa.gr} \\
}

\renewcommand{\headeright}{}
\renewcommand{\undertitle}{}

\hypersetup{
pdftitle={Anisotropic Conformal Dark Gravity on the Lorentz Tangent Bundle Spacetime},
pdfsubject={gr-qc},
pdfauthor={Christos Savvopoulos, Panayiotis Stavrinos},
pdfkeywords={dark matter, dark energy, anisotropic gravitational field, conformal gravity, tangent bundle, Finsler-like, generalized Friedman equations, modified gravity},
}

\begin{document}
\maketitle

\begin{abstract}
	In this work we investigate the anisotropic conformal structure of  the gravitational field incorporating dark gravity in a generalized Lagrange geometric framework on the Lorentz tangent bundle and we present two applications; the anisotropic conformal Minkowski spacetime and the anisotropic conformal FLRW cosmology. In the first application, the conformal factor induces an anisotropic conformal de-Sitter-like space with extra curvature which causes extra gravity and allows for Sasaki-type Finsler-like structures which could potentially describe certain gravitational phenomena in a more extended form. The cosmological properties of the model are also studied using a FLRW metric structure for the underlying base manifold in the second application, where we derive generalized Friedmann-like equations for the horizontal subspace of the Lorentz tangent bundle spacetime that reduce under certain conditions to those given by  A. Triantafyllopoulos and P. C. Stavrinos (2018) [Class. Quantum Grav. 35 085011] as well as those of general relativity.
\end{abstract}


\section{Introduction}
\label{intro}
\paragraph*{}Over the last decades the topic of dark matter and dark energy stands at the forefront of scientific research in the field of gravity and cosmology \cite{farnes2018unifying, nadler2020milky, ren2018constraining, dror2020directly, rigault2020strong,Cai:2009zp, Valentino_2020, refId0, leibundgut2018type, velten2018gauging, hoscheit2018kbc, di2020nonminimal,Faraoni:2014vra, collett2018precise, OKS21,Geng:2011ka, Farrah_22023, Farrah_2023}. The significant interest in this topic stems from observational data that attribute the vast majority of the mass in the observable universe to sources other than ordinary luminous matter, what researchers called dark matter \cite{hartle_2021}. Examples of phenomena that would suggest a modified theory of gravity that would account for the discrepancies in the classical theory of gravitation due to the presence of dark matter and dark energy arise from the study of gravitational lensing, cosmic microwave background radiation (CMB) or the rotational curves of spiral galaxies \cite{godlowski_2007, Persic:1995ru}.
\paragraph*{}The study of such phenomena suggests that dark matter would contribute significantly in the evolution and acceleration of the universe which would mean that the study of dark matter is essential for cosmology. In particular, dark matter could possibly be considered as the main reason for galaxy structure formations and dark energy as the drive for the measured cosmic acceleration \cite{DanieleBertacca_2011, Tsujikawa2011, Tsujikawa2006}. This would suggest the need for a modified theory of gravity that would incorporate such gravitational effects and potentially describe the aforementioned phenomena, since extra dark gravity influences all scales of matter. Particularly, the $\Lambda$-CDM model is especially efficient in agreeing with observational data \cite{Pan:2017zoh,Yang:2018qmz,Xu}. However, it has been argued \cite{Bouali2023} that this model is lacking in sufficient mathematical and theoretical background. This would therefore indicate that there exists a need to obtain an improved mathematical structure consistent with such description of the universe with dark gravity. 
\paragraph*{}There is some evidence that a conformal theory of gravity can dynamically accommodate for this "extra" gravity by introducing additional degrees of freedom to the existing underlying metric structures \cite{Chamseddine2013, Yadav_2019, YANG201343}. In addition, a conformally invariant theory could possibly be linked to a bounce evolution of the universe \cite{Boisseau_2016, Paul_2014}. Furthermore, the purely gravitational dark matter may be produced mainly by the gravitational particle creation process \cite{Yokoyama19}, which is thought to normally convert anisotropy energy into radiation energy \cite{Misner1973}.
It is also worth noting that a conformal framework of gravity seems to be particularly compatible with observational data of galactic rotational velocities and halos among others \cite{Nesbet13, Bertacca2010, Mannheim_2013, KeithHorne16}.
\paragraph*{}One geometrical frame for the anisotropic conformal modification of gravity arises from the extension of the underlying geometry of a manifold $(M,g(x))$; i.e. generalized Lagrange metric structures on the tangent bundle \cite{Ikedastav, Stav10, Stav12, Stav13, Stav19, Stav2013, MAVROMATOS11, Triantafyllopoulos_2018}. In this framework the gravitational field is extended in a higher dimensional space with greater volume. A Sasaki-type Finsler-like structure of this kind not only furnishes the geometric frame with extra degrees of freedom, but also endows the structure of the spacetime with local anisotropy and extra dimensions, which could be associated with dark gravitational effects \cite{Coimbra2013}, while simultaneously preserving the light cone \cite{Javaloyes2019, Javaloyes2020}. These extra degrees of freedom are introduced in 8-dimensions and are linked to the notion of direction-dependent anisotropy caused by velocity or momentum coordinates \cite{Ikedastav}. This dependence of the physical quantities on the observer 4-velocity provides a natural geometric extension of the Riemannian frame on the tangent bundle, which could be reproduced from the generalized frame by eliminating this direction dependence. Moreover, such a Sasaki-type consideration could potentially be related to a generalized anisotropic conformal de-Sitter Minkowski spacetime structure. We can notice that a Friedmann spacetime is isotropic conformal to a Minkowski flat spacetime. Analogously, it could be interesting to study an anisotropic conformal Minkowski as well as FLRW spacetime using the aforementioned geometry. Finally, due to the strong association of Finsler and Finsler-like geometries with the effective geometry within anisotropic media \cite{Ikedastav, born1999, Silva_2021},
forming a natural gravitational analogy \cite{Ikedastav, Barcel2005}, it could be argued that they could play an important role in quantum gravity considerations \cite{KOSTELECKY11, Harkoquantum, Vacaru11, Vacaru_2011}.
\paragraph*{}This work is organized as follows: in Section \ref{sec2}, we present the generalized Lagrange Sasaki-type geometric structure of the tangent bundle giving the relations for the metric, the connection, the curvature tensor field as well as the field equations, among others. In Section \ref{sec3}, we give the geodesic equations for this model. In Section \ref{secflat} we study the case of the anisotropic conformal Minkowski spacetime and derive a couple of special types of conformal factors. Further in section \ref{secFRW} we investigate the anisotropic conformal FLRW-cosmology using the geometric frameworks developed in this work. Finally, in section \ref{conclusion}, we summarize our results of and in Appendix A we present some further geometric results.
\section{Metric Structure}\label{sec2}
In this section, we shall introduce some basic notions from the geometry of generalized Finsler-like metric structures. Let $M$ be a differentiable manifold of dimension $\dim(M)=n$ and $TM$ be its tangent bundle. Let the manifold $M$ be endowed with a (pseudo-)Riemannian metric $\gamma(x)$. Then it is well known \cite{Sasaki58, Sasaki61} that its tangent bundle can be endowed with a Riemann-Sasaki metric structure as follows:
\begin{equation}
    dl^2=
\tensor{\gamma}{_{\mu\nu}}(x)dx^\mu \otimes dx^\nu+\tensor{\gamma}{_{ab}}(x)\delta y^a\otimes \delta y^b
\end{equation}
where
\begin{equation}
    \delta y^a=dy^a+\tensor{N}{^a_\mu} (x,y)dx^\mu
\end{equation}
with $\mu, \nu,\dots = 0, 1, \cdots, n-1$ and
$a, b, \dots = 0, \cdots, n-1$. The components of $\tensor{N}{^a_{\mu}}(x,y)$, which is known as the non-linear connection, is produced by the Whitney sum of the horizontal and vertical subspaces of the tangent bundle \cite{Miron1994, VacaruStav}. It is then well-established that this metric structure for the tangent bundle can be further generalized to include Finsler, Lagrange and generalized Lagrange metrics $g(x,y)$, collectively known as Finsler-like structures. In this case we have:
\begin{equation}
    d\tau^2=
\tensor{g}{_{\mu\nu}}(x,y)dx^\mu \otimes dx^\nu+\tensor{g}{_{ab}}(x,y)\delta y^a\otimes \delta y^b
\end{equation}

\paragraph*{}Let us now consider a non-reducible generalised Lagrange tangent bundle space $TM$, 
\begin{equation}  
GL^{(2n)}=\left(g_{\mu\nu}(x,y), g_{ab}(x,y)\right)
\end{equation}
with metric $\mathcal{G}$ such that
\begin{equation}\label{metric}
   ds^2= e^{f(x,y)}dl^2=\sigma(x,y)dl^2
\end{equation}
where $f,\sigma: TM\rightarrow \mathbb{R}$ are functions which are at least $C^2$ known as the (anisotropic) conformal factors. For convenience we shall be using both of these equivalent definitions for the conformal factor throughout this study. Physically, the conformal factor is introduced to incorporate the dark gravitational effect into the geometric framework of the gravitational field. The variable $y$ in particular, is the internal variable that introduces direction dependence and hence local anisotropy. If the conformal factor does not depend on $y$, this is interpreted as isotropic dark gravity, and if $f=0$ then we get a spacetime without dark gravity. This metric space is said to be anisotropic conformal \cite{Javaloyes2020, Moon17} to the Riemann-Sasaki metric space defined by the Riemannian metric $\gamma(x)$.
In terms of the bundle components, the metric tensor can be equivalently written as, 
\begin{equation}
    \tensor{\mathcal{G}}{_{MN}}(x,y)=\{\tensor{g}{_{\mu\nu}}(x,y), \tensor{g}{_{ab}}(x,y)\}
\end{equation}
where $M, N,\dots = 0, 1, \cdots, 2n-1$
\begin{eqnarray}
&&g_{\mu\nu}(x,y)=e^{f(x,y)}\gamma_{\mu\nu}(x)\\ \label{HVmetricrel}
    &&\tensor{g}{_{ab}}(x,y)=\tensor{\delta}{^\mu_a}\tensor{\delta}{^\nu_b}\tensor{g}{_{\mu\nu}}(x,y)
\end{eqnarray}
and $\gamma_{\mu\nu}(x)$ is a Riemannian metric that has been extended in the vertical subspace as $\tensor{\gamma}{_{ab}}=\tensor{\delta}{^\mu_a}\tensor{\delta}{^\nu_b}\tensor{\gamma}{_{\mu\nu}}$.
\paragraph*{}The adapted and dual bases of TTM are given by 
\begin{equation}
  X_M=\{\delta_\mu, \bar\p_a\}  ~,~ X^M=\{dx^\mu, \delta y^a \}  
\end{equation}
respectively, where 
\begin{equation}
\delta_\mu=\p_\mu-N^a_\mu(x,y)\bar\p_a~,~
\p_\mu := \frac{\p}{\p x^\mu}
~~,~~
\bar\p_a:= \frac{\p}{\p y^a}
\end{equation}
The connection is then given by the following:
\beqa
&&D_{\delta_\nu}\delta_\mu=
\tensor{L}{^\lambda_{\mu\nu}}\delta_\lambda~~~,~~~
D_{\delta_\nu}\bar\p_a=
\tensor{\tilde L}{^c_{a\nu}}\bar\p_c\nn \\
&&D_{\bar\p_b}\delta_\mu=
\tensor{\tilde C}{^\lambda_{\mu b}}\delta_\lambda~~~,~~~
D_{\bar\p_b}\bar\p_a=
\tensor{C}{^c_{ab}}\bar\p_c
\eeqa
Hence, the coefficients of the d-connection are
\beqa
\tensor{\bm\Gamma}{^L_{MN}}=\{
\tensor{L}{^\lambda_{\mu\nu}}, \tensor{\tilde L}{^c_{a\nu}}, \tensor{\tilde C}{^\lambda_{\mu b}}, 
\tensor{C}{^c_{ab}}
\}
\eeqa
The d-connection preserves the horizontal and vertical components of a vector under parallel translation \cite{Miron1994}. Throughout this study we shall assume a metrical d-connection \cite{Miron1994, MWI, VacaruStav, Vacaru2Stav}. From this assumption we get the subsequent relations for the d-connection coefficients:
\begin{eqnarray}\label{d1}
    &&\tensor {L}{^\lambda_{\mu\nu}}
=\tensor{\gamma}{^\lambda_{\mu \nu}}
+\frac{
\tensor{\delta}{^\lambda_{\nu}}\delta_{\mu}\sigma+
\tensor{\delta}{^\lambda_ \mu} \delta_\nu \sigma-\gamma_{\mu\nu}\gamma^{\lambda\rho}\delta_\rho \sigma}{2\sigma}\\ \label{d2}
&&\tensor{\tilde L}{^a_{b\mu}}=\frac{\bar\p _b \tensor{N}{^a_\mu}+\tensor{\gamma}{^{ac}}\p_\mu\tensor{\gamma}{_{bc}}+\tensor{\delta}{^a _b}\frac{\delta_\mu\sigma}{\sigma} -\tensor{\gamma}{_{bd}}\tensor{\gamma}{^{ac}}\bar\p _c\tensor{N}{^d _\mu}}{2}\\\label{d3}
&&\tensor {C}{^c_{ab}}
=\frac{1}{2\sigma}\left(
\tensor{\delta}{^c_{b}}\bar\p_{a}\sigma+
\tensor{\delta}{^c_{a}}\bar\p_{b}\sigma-\gamma_{ab }\gamma^{c d}\bar\p_{d}\sigma
\right)\\ &&\label{d4}
\tensor {\tilde C}{^\lambda_{\mu c}}=\frac{1}{2}\tensor{\delta}{^\lambda_\mu}\bar\p_c\left(\ln\sigma\right)
\end{eqnarray}
where $\tensor{\gamma}{^\lambda_{\mu \nu}}$ are the Christoffel symbols of the Riemannian metric $\gamma$.
\paragraph*{}Using the d-connection coefficients we previously found in relations (\ref{d1}-\ref{d4}), we can now calculate the curvature of this space. In particular, let $\mathcal{R}$ be the curvature tensor field of the d-connection $D$, then the non-zero components of $\mathcal{R}$ are given by the following relations: 
\begin{eqnarray}
    &&\tensor{\mathcal{R}}{_\nu ^{\mu}_{ \rho \sigma}}=\tensor{R}{_\nu ^{\mu}_{ \rho \sigma}} ~~,~~
     \tensor{\mathcal{R}}{_b ^{a}_{ \kappa \lambda}}=\tensor{R}{_b ^{a}_{ \kappa \lambda}}\\
     &&\tensor{\mathcal{R}}{_\nu ^{\mu}_{ \rho d}}=\tensor{P}{_\nu ^{\mu}_{ \rho d}}~~,~~
     \tensor{\mathcal{R}}{_b ^{a}_{ \kappa d}}=\tensor{P}{_b ^{a}_{ \kappa d}}\\
     &&\tensor{\mathcal{R}}{_\nu ^{\mu}_{ a b}}=\tensor{S}{_\nu ^{\mu}_{ a b}} ~~,~~
     \tensor{\mathcal{R}}{_b ^{a}_{ c d}}=\tensor{S}{_b ^{a}_{ c d}}
\end{eqnarray}
where the d-tensor fields are given by:
\begin{eqnarray}
    &&\tensor{R}{_\nu ^{\mu}_{ \rho \sigma}}=\delta_\sigma \tensor{L}{^\mu _{\nu \rho}}-\delta_\rho \tensor{L}{^\mu_{\nu \sigma}}+\tensor{L}{^\kappa_{\nu \rho}}\tensor{L}{^\mu _{\kappa \sigma}}-\tensor{L}{^\kappa_{\nu \sigma}}\tensor{L}{^\mu _{\kappa \rho}}+\tensor{\tilde{C}}{^\mu _{\nu c}}\tensor{R}{^c _{\rho \sigma}}\\
    &&\tensor{R}{_b ^{a}_{ \rho \sigma}}=\delta_\sigma \tensor{\tilde{L}}{^a _{b \rho}}-\delta_\rho \tensor{\tilde{L}}{^a_{b \sigma}}+\tensor{\tilde{L}}{^c_{b \rho}}\tensor{\tilde{L}}{^a _{c \sigma}}-\tensor{\tilde{L}}{^c_{b \sigma}}\tensor{\tilde{L}}{^a _{c \rho}}+\tensor{C}{^a _{b c}}\tensor{R}{^c _{\rho \sigma}} \\
    &&\tensor{P}{_\nu ^{\mu}_{ \rho d}}=\bar\p_d\tensor{L}{^\mu_{\nu\rho}}-\tensor{\tilde{C}}{^\mu _{\nu d|\rho}}+\tensor{\tilde{C}}{^\mu_{\nu b}}\tensor{P}{^b_{\rho d}}\\
    &&\tensor{P}{_b ^{a}_{ \rho d}}=\bar\p_d\tensor{\tilde{L}}{^a _{b\rho}}-\tensor{C}{^a _{b d|\rho}}+\tensor{C}{^a_{b c}}\tensor{P}{^c_{\rho d}}\\
    &&\tensor{S}{_\nu ^{\mu}_{ a b}}=\bar\p_b \tensor{\Tilde{C}}{^\mu _{\nu a}}-\bar\p_a \tensor{\Tilde{C}}{^\mu _{\nu b}}+\tensor{\tilde{C}}{^\lambda _{\nu a}}\tensor{\tilde{C}}{^\mu _{\lambda b}}-\tensor{\tilde{C}}{^\lambda _{\nu b}}\tensor{\tilde{C}}{^\mu _{\lambda a}}\\
    &&\tensor{S}{_b ^{a}_{ c d}}=\bar\p_d \tensor{C}{^a _{b c}}-\bar\p_c \tensor{C}{^a _{b d}}+\tensor{C}{^e _{bc}}\tensor{C}{^a _{ed}}-\tensor{C}{^e _{bd}}\tensor{C}{^a _{ec}}
\end{eqnarray}
where $\tensor{R}{^c _{\rho \sigma}}=\delta_\sigma \tensor{N}{^c_\rho}-\delta_\rho \tensor{N}{^c_\sigma}=\tensor{\delta}{_{[\sigma}} \tensor{N}{^c_{\rho]}}$. $N$ is said to be integrable if and only if $\tensor{R}{^c _{\rho \sigma}}=0$ \cite{Miron1994, MWI}. Let the non-linear connection be of Cartan-type, i.e. $\tensor{N}{^a_\kappa}=\tensor{\gamma}{^a_{b\kappa }}y^b$. Then it is clear that in general $N$ is not integrable. For the anisotropic conformal metric (\ref{metric}) we have the following curvature tensor field components\footnote{The study of the $P-$curvature lies outside the scope of the present study and shall be henceforth omitted.}:
\begin{eqnarray}
   &&\tensor{R}{_\nu ^{\mu}_{ \rho \sigma}}=\tensor{K}{_\nu ^{\mu}_{ \rho \sigma}}+\frac{1}{2}\tensor{\mathcal{L}}{_\nu ^{\mu}_{ \rho \sigma}}+ \frac{1}{4}\tensor{M}{_\nu ^{\mu}_{ \rho \sigma}}\\
   &&\tensor{R}{_b ^{a}_{ \rho \sigma}}=\frac{1}{2}\tensor{\tilde{\mathcal{L}}}{_b ^{a}_{ \rho \sigma}}+\frac{1}{4}\tensor{\tilde{M}}{_b ^{a}_{ \rho \sigma}}\\
   &&\tensor{S}{_\nu ^{\mu}_{ a b}}=0\\
   &&\tensor{S}{_b ^{a}_{ c d}}=\frac{1}{2}\big(
\tensor{\delta}{^a_{[c}}\tensor{\bar\p}{_{d]}}\bar\p_{b}f
+\tensor{\gamma}{_{b[d}}\gamma^{a e}\tensor{\bar\p}{_{c]}}\bar\p_{e}f\big) +\frac{1}{4}\big(
\tensor{\delta}{^a_{[d}}\bar\p_{b}f\tensor{\bar\p}{_{c]}}f+\tensor{\delta}{^a_{[c}}\tensor{\gamma}{_{d]b}}\gamma^{e f}\bar\p_{e}f\bar\p_{f}f+\tensor{\gamma}{_{b[c}}\gamma^{a e}\tensor{\bar\p}{_{d]}}f\bar\p_{e}f\big) \label{Scurvtens}
\end{eqnarray}
where 
\begin{eqnarray}
   &&\tensor{K}{_\nu ^{\mu}_{ \rho \sigma}}=\tensor{\p}{_{[\sigma}}\tensor{\gamma}{^\mu _{\rho]\nu}}+\tensor{\gamma}{^\kappa _{\nu[\rho}}\tensor{\gamma}{^\mu _{\sigma]\kappa}}\\
   &&\tensor{\mathcal{L}}{_\nu ^{\mu}_{ \rho \sigma}}=\tensor{\delta}{^\mu_{[\rho}}\tensor{\delta}{_{\sigma]}}\tensor{\delta}{_\nu}f + \tensor{\delta}{^\mu_{\nu}}\tensor{\delta}{_{[\sigma}}\tensor{\delta}{_{\rho]}}f + \tensor{\p}{_{[\rho}}\tensor{\gamma}{_{\sigma]\nu}}\tensor{\gamma}{^{\mu\lambda}}\tensor{\delta}{_\lambda}f +\tensor{\gamma}{_{\nu[\sigma}}\tensor{\p}{_{\rho]}}\tensor{\gamma}{^{\mu\lambda}}\tensor{\delta}{_\lambda}f + \tensor{\gamma}{_{\nu[\sigma}}\tensor{\gamma}{^{\mu\lambda}}\tensor{\delta}{_{\rho]}}\tensor{\delta}{_\lambda}f \nn \\
   &&+ \tensor{\gamma}{^\kappa _{\nu[\rho}}\tensor{\delta}{^\mu _{\sigma]}}\tensor{\delta}{_\kappa}f + \tensor{\gamma}{^\kappa _{\nu[\sigma}}\tensor{\gamma}{_{\rho]\kappa}}\tensor{\gamma}{^{\mu\lambda}}\tensor{\delta}{_\lambda}f + \tensor{\gamma}{^\mu _{\kappa[\rho}}\tensor{\gamma}{_{\sigma]\nu}}\tensor{\gamma}{^{\kappa \lambda}}\tensor{\delta}{_\lambda}f+ \tensor{\delta}{^\mu_\nu}\tensor{\delta}{_{[\sigma}}\tensor{N}{^c_{\rho]}}\bar\p_c f\\
   &&\tensor{M}{_\nu ^{\mu}_{ \rho \sigma}}=\tensor{\delta}{^\mu_{[\sigma}}\tensor{\delta}{_{\rho]}}f\tensor{\delta}{_\nu}f +\tensor{\gamma}{ _{\nu[\rho}}\tensor{\gamma}{^{\mu\lambda}}\tensor{\delta}{ _{\sigma]}}f\tensor{\delta}{_\lambda}f+ \tensor{\delta}{^\mu_{[\rho}}\tensor{\gamma}{_{\sigma]\nu}}\tensor{\gamma}{^{\kappa \lambda}}\tensor{\delta}{_\kappa}f\tensor{\delta}{_\lambda}f\\
   &&\tensor{\tilde{\mathcal{L}}}{_b ^{a}_{ \rho \sigma}}= \tensor{\delta}{_{[\sigma}}\bar\p _b \tensor{N}{^a_{\rho]}}+\tensor{\p}{_{[\sigma}}\tensor{\gamma}{^{ac}}\tensor{\p}{_{\rho]}}\tensor{\gamma}{_{bc}}+\tensor{\delta}{^a _b}\tensor{\delta}{_{[\sigma}}\tensor{\delta}{_{\rho]}}f +\tensor{\gamma}{^{ac}}\tensor{\p}{_{[\rho}}\tensor{\gamma}{_{bd}}\bar\p _c\tensor{N}{^d _{\sigma]}} +\tensor{\gamma}{_{bd}}\tensor{\p}{_{[\rho}}\tensor{\gamma}{^{ac}}\bar\p _c\tensor{N}{^d _{\sigma]}}\nn \\
    &&+\tensor{\gamma}{_{bd}}\tensor{\gamma}{^{ac}}\tensor{\delta}{_{[\rho}}\bar\p _c\tensor{N}{^d _{\sigma]}}+\tensor{\delta}{_{[\sigma}} \tensor{N}{^a_{\rho]}}\bar\p_{b}f+\tensor{\delta}{_{[\sigma}} \tensor{N}{^c_{\rho]}}\tensor{\delta}{^a_{b}}\bar\p_{c}f-\tensor{\delta}{_{[\sigma}} \tensor{N}{^c_{\rho]}}\gamma_{bc}\gamma^{a d}\bar\p_{d}f\\
    &&\tensor{\tilde{M}}{_b ^{a}_{ \rho \sigma}}=\bar\p _b \tensor{N}{^c_{[\rho}}\bar\p _c \tensor{N}{^a_{\sigma]}}+\tensor{\gamma}{^{cd}}\bar\p _c \tensor{N}{^a_{[\sigma}}\tensor{\p}{_{\rho]}}\tensor{\gamma}{_{bd}}+\tensor{\gamma}{^{ad}}\bar\p _b \tensor{N}{^c_{[\rho}}\tensor{\p}{_{\sigma]}}\tensor{\gamma}{_{cd}}+\tensor{\gamma}{_{bd}}\tensor{\gamma}{^{ce}}\bar\p _e\tensor{N}{^d _{[\sigma}}\bar\p _c \tensor{N}{^a_{\rho]}} \nn \\
   &&+\tensor{\gamma}{^{ae}}\tensor{\gamma}{^{cd}}\tensor{\p}{_{[\sigma}}\tensor{\gamma}{_{ce}}\tensor{\p}{_{\rho]}}\tensor{\gamma}{_{bd}}+\tensor{\gamma}{_{bd}}\tensor{\gamma}{^{af}}\tensor{\gamma}{^{ce}}\bar\p _e\tensor{N}{^d _{[\sigma}}\tensor{\p}{_{\rho]}}\tensor{\gamma}{_{cf}} +\tensor{\gamma}{_{bd}}\tensor{\gamma}{^{ac}}\bar\p _c\tensor{N}{^d _{[\sigma}}\tensor{\delta}{_{\rho]}}f+\tensor{\gamma}{_{cd}}\tensor{\gamma}{^{ae}}\bar\p _e\tensor{N}{^d _{[\rho}}\bar\p _b \tensor{N}{^c_{\sigma]}}\nn \\
   &&+\tensor{\gamma}{^{ac}}\bar\p _c\tensor{N}{^d _{[\rho}}\tensor{\p}{_{\sigma]}}\tensor{\gamma}{_{bd}}+\tensor{\gamma}{_{bc}}\tensor{\gamma}{^{ad}}\bar\p _d\tensor{N}{^c _{[\rho}}\tensor{\delta}{_{\sigma]}}f +\tensor{\gamma}{_{bd}}\tensor{\gamma}{^{ae}}\bar\p _c\tensor{N}{^d _{[\rho}}\bar\p _e\tensor{N}{^c _{\sigma]}}
\end{eqnarray}
$K$, in particular, is the Riemann curvature tensor corresponding to the (pseudo-)Riemannian metric $\gamma$ of the underlying manifold structure. The horizontal $R$-curvature contains extra terms, in addition to the underlying Riemannian $K$-curvature, which allow for any discrepancies to the Riemannian $K$-curvature that result from the effect of the "extra" dark gravity and could otherwise be interepreted as perturbations to the Riemannian framework to be incorporated in the geometry of the spacetime in the tangent bundle. 
\paragraph*{}A physical interpretation of the vertical $S$-curvature (\ref{Scurvtens}), on the other hand, could be tied to an anisotropic behavior of dark gravity since the $S$-curvature indicates an anisotropically curved spacetime. This is made evident by the above-mentioned form of the vertical $S$-curvature which depends on the existence of a direction-dependent conformal factor, which in turn presupposes an anisotropic dark matter as mentioned in the beginning. A non-trivial $S$-curvature is absent from a Riemannian framework and would thus introduce extra degrees of freedom not present in a Riemannian  theory of gravity. Consequently, this geometric structure could allow for a broader study of gravitational phenomena linked with dark gravity and anisotropy (e.g. the evolution of universe).
\paragraph*{}We shall now find the Ricci tensors as follows:
\begin{eqnarray}\label{Riccigentens}
   \tensor{R}{_{\nu\rho}}=\tensor{R}{_\nu ^{\mu}_{ \rho \mu}}=\tensor{K}{_{\nu \rho}}+\frac{1}{2}\tensor{\mathcal{L}}{_{\nu\rho}}+\frac{1}{4}\tensor{M}{_{\nu\rho}}
   \end{eqnarray}
   where
   \begin{eqnarray}
   \tensor{\mathcal{L}}{_{\nu\rho}}=\tensor{\delta}{_{[\nu}}\tensor{\delta}{_{\rho]}}f + \tensor{\gamma}{_{\nu[\mu}}\tensor{\gamma}{^{\mu\lambda}}\tensor{\delta}{_{\rho]}}\tensor{\delta}{_\lambda}f+ \tensor{\p}{_{[\rho}}\tensor{\gamma}{_{\mu]\nu}}\tensor{\gamma}{^{\mu\lambda}}\tensor{\delta}{_\lambda}f +\tensor{\gamma}{_{\nu[\mu}}\tensor{\p}{_{\rho]}}\tensor{\gamma}{^{\mu\lambda}}\tensor{\delta}{_\lambda}f + \tensor{\gamma}{^\kappa _{\nu[\mu}}\tensor{\gamma}{_{\rho]\kappa}}\tensor{\gamma}{^{\mu\lambda}}\tensor{\delta}{_\lambda}f + \tensor{\gamma}{^\mu _{\kappa[\rho}}\tensor{\gamma}{_{\mu]\nu}}\tensor{\gamma}{^{\kappa \lambda}}\tensor{\delta}{_\lambda}f+\tensor{\delta}{_{[\nu}}\tensor{N}{^c_{\rho]}}\bar\p_c f
   \end{eqnarray}
   and 
   \begin{eqnarray}
   \tensor{M}{_{\nu\rho}}= \tensor{\gamma}{ _{\nu[\rho}}\tensor{\gamma}{^{\mu\lambda}}\tensor{\delta}{ _{\mu]}}f\tensor{\delta}{_\lambda}f
\end{eqnarray}
and the Ricci tensor corresponding to the Riemannian metric $\gamma$ is
\begin{equation}
    \tensor{K}{_{\nu \rho }}=\tensor{K}{_\nu ^{\mu}_{ \rho \mu}}=\tensor{\p}{_{[\mu}}\tensor{\gamma}{^\mu _{\rho]\nu}}+\tensor{\gamma}{^\kappa _{\nu[\rho}}\tensor{\gamma}{^\mu _{\mu]\kappa}}
\end{equation}
From the $S$-curvature, we get:
\begin{equation}
    \tensor{S}{_{bc}}=\tensor{S}{_b ^{a}_{ c a}}=\frac{1}{4}\bigg(2\tensor{\gamma}{_{b[a}}\gamma^{a d}\tensor{\bar\p}{_{c]}}\bar\p_{d}f+\tensor{\gamma}{_{b[c}}\gamma^{a d}\tensor{\bar\p}{_{a]}}f\bar\p_{d}f\bigg)
\end{equation}
We therefore have the following scalar curvature:
\begin{equation}\label{ricciscalargeneral}
    \mathcal{R}=R+S=e^{-f}\bigg(K+\frac{1}{2}\mathcal{L}+\frac{1}{4}M+\frac{3}{4}\tilde{S}\bigg)
\end{equation}
where
\begin{eqnarray}
   R=\tensor{g}{^{\nu \rho}}\tensor{R}{_{\nu\rho}}=e^{-f}\bigg(K+\frac{1}{2}\mathcal{L}+\frac{1}{4}M\bigg)
\end{eqnarray} 
with
\begin{eqnarray}
\mathcal{L}&=&-(n-1)\tensor{\gamma}{^{\mu\lambda}}\tensor{\delta}{_{\mu}}\tensor{\delta}{_\lambda}f+ \tensor{\gamma}{^{\nu \rho}}\tensor{\p}{_{[\rho}}\tensor{\gamma}{_{\mu]\nu}}\tensor{\gamma}{^{\mu\lambda}}\tensor{\delta}{_\lambda}f -(n-1)\tensor{\p}{_{\mu}}\tensor{\gamma}{^{\mu\lambda}}\tensor{\delta}{_\lambda}f+ \tensor{\gamma}{^{\nu \rho}}\tensor{\gamma}{^\kappa _{\nu[\mu}}\tensor{\gamma}{_{\rho]\kappa}}\tensor{\gamma}{^{\mu\lambda}}\tensor{\delta}{_\lambda}f \nn \\
   && -(n-1) \tensor{\gamma}{^\mu _{\kappa\mu}}\tensor{\gamma}{^{\kappa \lambda}}\tensor{\delta}{_\lambda}f +\tensor{\gamma}{^{\nu \rho}}\tensor{\delta}{_{[\nu}}\tensor{N}{^c_{\rho]}}\bar\p_c f \\
   M&=& \frac{e^{-f}}{4}(n-1)\tensor{\gamma}{^{\mu\lambda}}\tensor{\delta}{ _{\mu}}f\tensor{\delta}{_\lambda}f
\end{eqnarray}
and the Ricci scalar corresponding to the Riemannian metric $\gamma$ is
\begin{equation}
    K=\tensor{\gamma}{^{\nu \rho}}\tensor{K}{_{\nu \rho }}
\end{equation}
For the scalar $S$-curvature we have
\begin{equation}
    S=\tensor{g}{^{b c}}\tensor{S}{_{bc}}=\frac{3}{4}e^{-f}\tilde{S}
\end{equation}
with
\begin{equation}    
    \tilde{S}=\frac{1}{3}\bigg(-2(n-1)\tensor{\gamma}{^{a b}}\tensor{\bar\p}{_{a}}\tensor{\bar\p}{_{b}}f+(n-1)\gamma^{a b}\tensor{\bar\p}{_{a}}f\tensor{\bar\p}{_{b}}f\bigg)
\end{equation}
The scalar $S$-curvature could be interpreted as the degree of anisotropy of a conformal anisotropically curved spacetime which includes anisotropic gravitational effects as shown in the previous relation. It is worth pointing out, however, that while the $S$-curvature is dominated by the direction dependence of the conformal factor, a flat vertical space does not necessarily lead to (or result from) a direction independent conformal factor and such a case should be treated carefully as shall be demonstrated in a later section of this study.
\paragraph*{} The field equations are then given by the calculus of variation on the action given in \cite{MWI, Miron1994, Triantafyllopoulos_2018} \footnote{The gravitational constant has been taken as $1$.}:
\beqa\label{fieldzero}
\tensor{\mathcal{R}}{_{MN}}-\frac{1}{2}\mathcal{R}\tensor{\mathcal{G}}{_{MN}}= \tensor{\mathcal{T}}{_{MN}}
\eeqa
where $\tensor{\mathcal{T}}{_{MN}}$ is the energy-momentum tensor field on the tangent bundle in the adapted basis; namely $\tensor{\mathcal{T}}{_{\mu\nu}}=\tensor{T}{_{\mu\nu}}$ and $\tensor{\mathcal{T}}{_{ab}}=\tensor{W}{_{ab}}$ with $\tensor{\mathcal{T}}{_{a\nu}}=\tensor{\mathcal{T}}{_{\mu b}}=0$.
By taking the trace of relation (\ref{fieldzero}) we get the following:
\begin{equation}
    \mathcal{R}=-\frac{1}{n-1}\mathcal{T}
\end{equation}
where $\mathcal{T}=\tensor{\mathcal{G}}{^{MN}}\tensor{\mathcal{T}}{_{MN}}$ or, equivalently, $\mathcal{T}=T+W$, where $T=\tensor{g}{^{\mu\nu}}\tensor{T}{_{\mu\nu}}$ is the trace of the horizontal component of the energy-momentum tensor and $W=\tensor{g}{^{ab}}\tensor{W}{_{ab}}$ is the trace of the vertical component of the energy-momentum tensor, respectively. We thus arrive at the following equivalent form of the field equations:
\begin{equation} \label{fieldeq}
\tensor{\mathcal{R}}{_{MN}}=\tensor{\mathcal{T}}{_{MN}}-\frac{1}{2n-2}\mathcal{T}\tensor{\mathcal{G}}{_{MN}}
\end{equation}
By virtue of relations (\ref{HVmetricrel}), (\ref{fieldzero}) and (\ref{fieldeq}) the following proposition holds: the horizontal and vertical Ricci curvatures are equal to each other if and only if the horizontal and vertical components of the energy-momentum tensors are also equal to each other\footnote{Unless otherwise stated explicitly, we shall henceforth assume that the Ricci curvatures are not equal to each other.}; namely:
\begin{equation}\label{prop}
    \tensor{W}{_{ab}}=\tensor{\delta}{^\mu _a}\tensor{\delta}{^\nu _b}\tensor{T}{_{\mu\nu}} \Longleftrightarrow \tensor{S}{_{ab}}=\tensor{\delta}{^\mu _a}\tensor{\delta}{^\nu _b}\tensor{R}{_{\mu\nu}}
\end{equation}
As can be seen from relation (\ref{prop}), an intrinsically geometric connection of the horizontal and vertical subspaces results in a physical connection of the energy-momentum tensors and vice-versa. This is, however, also a consequence of the profound relation between the metric of the horizontal and vertical subspaces that has been assumed in relation (\ref{HVmetricrel}), without which relation (\ref{prop}) would not be true. 

\section{Geodesics}\label{sec3}
We define the absolute energy $\mathcal{E}$ as follows \cite{Miron1994}:
\begin{equation}
    \mathcal{E}:=\tensor{g}{_{ab}}y^ay^b=\sigma(x,y)\tensor{\gamma}{_{ab}}(x)y^ay^b
\end{equation}
Using the absolute energy we define the following tensor:
\begin{equation}
    \tensor{\Tilde{g}}{_{cd}}:=\frac{1}{2}\bar\p_c\bar\p_d\mathcal{E}=\frac{\mathcal{E}}{2\sigma }\sigma_{cd} +\tensor{\gamma}{_{ad}}y^a\sigma_c+\tensor{\gamma}{_{ac}}y^a\sigma_d+\tensor{\gamma}{_{cd}}\sigma
\end{equation}
where $\sigma_a=\bar\p_a\sigma$. We shall now calculate the inverse tensor $\Tilde{g}$ as follows:
\begin{equation}
    \tensor{\Tilde{g}}{^{ab}}=\tensor{g}{^{ac}}\tensor{g}{^{db}}\tensor{\Tilde{g}}{_{cd}}=\frac{\mathcal{E}}{2\sigma^3}\sigma^{ab} +y^a\frac{\sigma^b}{\sigma^2}+y^b\frac{\sigma^a}{\sigma^2}+\frac{\tensor{\gamma}{^{ab}}}{\sigma}
\end{equation}
We further define:
\begin{equation}
    G^a:=\frac{1}{4}\tensor{\Tilde{g}}{^{ab}}(y^k\bar\p_b\p_k\mathcal{E}-\p_b\mathcal{E})
\end{equation}
where $y^k=\tensor{\delta}{^i_a}y^a$. In our case:
\begin{eqnarray}
    &&G^a(x,y):=\frac{1}{4\sigma}\bigg(\frac{\mathcal{E}}{2}\sigma^{ab} +y^a\sigma^b+y^b\sigma^a+\tensor{\gamma}{^{ab}}\bigg)\times\nn\\
    &&\bigg(\sigma_b\p_\mu \tensor{\gamma}{_{cd}} y^cy^d y^\mu +2\sigma \p_\mu \tensor{\gamma}{_{bc}} y^cy^\mu +2\sigma_\mu \tensor{\gamma}{_{bc}}y^cy^\mu +\frac{\mathcal{E}}{\sigma}\sigma_{\mu b}y^\mu -\tensor{\delta}{^\mu _b}\sigma \p_\mu\tensor{\gamma}{_{cd}}y^cy^d+\tensor{\delta}{^\mu _b}\frac{\mathcal{E}}{\sigma}\sigma_\mu\bigg)
\end{eqnarray}
where $\sigma_\mu=\p_\mu\sigma$ and $\sigma^a=\tensor{g}{^{ab}}\bar\p_b\sigma$. We can now write the geodesic equation as follows:
\begin{equation}
   \frac{dy^a}{dt}+2G^a(x,y)=0, \,\,\,\, y^a=\frac{dx^a}{dt}
\end{equation}
We observe that:
\begin{equation}
    G^a(x,y)=\frac{1}{2} \tensor{\gamma}{^a_{\mu \nu}}y^\mu y^\nu + \frac{1}{2}r^a+\frac{1}{2}l^a
\end{equation}
where $\tensor{\gamma}{^a _{\mu \nu}}$ are the Christoffel symbols of second type of the Riemannian metric $\gamma$, 
\begin{equation}
    r^a=\frac{1}{2\sigma}\bigg(2\sigma_\mu y^ay^\mu-\frac{\mathcal{E}}{\sigma}\tensor{\gamma}{^{a\beta}}\sigma_\beta\bigg)
\end{equation}
is the conformal part of the geodesics corresponding to the Riemannian case, and 
\begin{eqnarray}
    &&l^a(x,y):=\frac{1}{2}\bigg(\frac{\mathcal{E}}{2\sigma}\sigma^{ab} +y^a\frac{\sigma^b}{\sigma}+y^b\frac{\sigma^a}{\sigma}\bigg)\times\nn\\
    &&\bigg(\sigma\tensor{\gamma}{_{bc\mu}}y^cy^\mu+\sigma_b\p_\mu \tensor{\gamma}{_{cd}} y^cy^d y^\mu +2\sigma_\mu \tensor{\gamma}{_{bc}}y^cy^\mu +\frac{\mathcal{E}}{\sigma}\sigma_{\mu b}y^\mu +\tensor{\delta}{^\mu _b}\frac{\mathcal{E}}{\sigma}\sigma_\mu\bigg)+\frac{\sigma^a}{\sigma}\p_\mu\tensor{\gamma}{_{cd}}y^cy^dy^\mu+\frac{\mathcal{E}}{\sigma}\tensor{\sigma}{_\mu ^a}y^\mu \nn\\
\end{eqnarray}
is the generalized Lagrange conformal part of the geodesics, where $\tensor{\gamma}{_{a \mu \nu}}$ are the Christoffel symbols of first type of the Riemannian metric $\gamma$. Therefore we can write the geodesic equation as follows:
\begin{equation}\label{eq57}
   \frac{dy^a}{dt}+\tensor{\gamma}{^a_{\mu \nu}}y^\mu y^\nu + r^a+l^a=0, \,\,\,\, y^a=\frac{dx^a}{dt}
\end{equation}
It is then clear, that the direction dependence introduced by $\sigma$ contributes through $l^a$ in the geodesics. If the conformal map $\sigma$ is independent of the direction variable $y$, i.e. $\sigma=\sigma(x)$, then $l^a=0$ and the geodesics reduce to their Riemannian counterpart. The internal $y$ coordinates of the gravitational field express the anisotropic dark structure through the function $\sigma(x,y)$. The additional terms $r^a(x,y)$ and $l^a(x,y)$ in rel. (\ref{eq57}) provide an anisotropic conformal type of metric geodesics that incorporate dark gravitational effects which is imprinted in the structure of this spacetime. Dark gravity plays an essential role in these perturbed conformal geodesics. 

\section{ Anisotropic Conformal Minkowski Spacetime}\label{secflat}
\paragraph*{}In this section we shall examine a first application of this geometric framework by using a Minkowski metric structure for the underlying manifold. This could be especially interesting for the cosmology of a post-inflation universe which evolves towards flatness, since this metrical model could be considered connected to an anisotropic generalization of a de-Sitter metric spacetime in which the scale factor causes a conformal structure for the spatial metric, e.g. in a Friedmann metric space.
In particular, let 
\begin{equation}
    \tensor{g}{_{\mu \nu}}(x,y)=e^{f(x,y)}\tensor{\eta}{_{\mu\nu}}
\end{equation}
where
\beqa
\tensor{\eta}{_{\mu\nu}}=\diag(1,-1,-1,-1)
\eeqa
for some $f:TM\rightarrow\mathbb{R}$ which is at least $C^2$. Let the non-linear connection be of Cartan-type, i.e. $\tensor{N}{^a_\kappa}=\tensor{\gamma}{^a_{b\kappa }}y^b$. Then the non-linear connection is obviously zero and the adapted basis $\{\delta_\mu, \bar\p_a\}$ coincides with the ordinary basis $\{\p_\mu, \bar\p_a\}$.
\paragraph*{}Since the curvature tensor of the underlying structure, $K=0$, it is relatively easy to find the curvature tensor $\mathcal{R}$. Indeed:
\begin{eqnarray}\label{confflatcurvtr}
   &&\tensor{R}{_\nu ^{\mu}_{ \rho \sigma}}=\frac{1}{2}\big( \tensor{\delta}{^\mu_{[\rho}}\tensor{\p}{_{\sigma]}}\tensor{\p}{_\nu}f + \tensor{\eta}{_{\nu[\sigma}}\tensor{\eta}{^{\mu\lambda}}\tensor{\p}{_{\rho]}}\tensor{\p}{_\lambda}f \big)  + \frac{1}{4}\big( \tensor{\delta}{^\mu_{[\sigma}}\tensor{\p}{_{\rho]}}f\tensor{\p}{_\nu}f +\tensor{\eta}{ _{\nu[\rho}}\tensor{\eta}{^{\mu\lambda}}\tensor{\p}{ _{\sigma]}}f\tensor{\p}{_\lambda}f + \tensor{\delta}{^\mu_{[\rho}}\tensor{\eta}{_{\sigma]\nu}}\tensor{\eta}{^{\kappa \lambda}}\tensor{\p}{_\kappa}f\tensor{\p}{_\lambda}f\big)\\
    &&\tensor{R}{_b ^{a}_{ \rho \sigma}}=0 ~~,~~
    \tensor{S}{_\nu ^{\mu}_{ a b}}=0 \\ \label{confflatcurvts}
    &&\tensor{S}{_b ^{a}_{ c d}}=\frac{1}{2}\bigg(
\tensor{\delta}{^a_{[c}}\tensor{\bar\p}{_{d]}}\bar\p_{b}f
+\tensor{\eta}{_{b[d}}\eta^{a e}\tensor{\bar\p}{_{c]}}\bar\p_{e}f\bigg)+\frac{1}{4}\bigg(
\tensor{\delta}{^a_{[d}}\bar\p_{b}f\tensor{\bar\p}{_{c]}}f+\tensor{\delta}{^a_{[c}}\tensor{\eta}{_{d]b}}\eta^{e f}\bar\p_{e}f\bar\p_{f}f +\tensor{\eta}{_{b[c}}\eta^{a e}\tensor{\bar\p}{_{d]}}f\bar\p_{e}f\bigg)
\end{eqnarray}
The Ricci tensors and scalars are then:
\begin{eqnarray}\label{confflatcurvR}
   &&\tensor{R}{_{\nu\rho}}=\frac{1}{4} \bigg(2\tensor{\eta}{_{\nu[\mu}}\tensor{\eta}{^{\mu\lambda}}\tensor{\p}{_{\rho]}}\tensor{\p}{_\lambda}f  + \tensor{\eta}{ _{\nu[\rho}}\tensor{\eta}{^{\mu\lambda}}\tensor{\p}{ _{\mu]}}f\tensor{\p}{_\lambda}f \bigg) \\ \label{confflatcurvS}
   &&\tensor{S}{_{bc}}=\frac{1}{4}\bigg(2\tensor{\eta}{_{b[a}}\eta^{a d}\tensor{\bar\p}{_{c]}}\bar\p_{d}f+\tensor{\eta}{_{b[c}}\eta^{a d}\tensor{\bar\p}{_{a]}}f\bar\p_{d}f\bigg)\\\label{confflatcurvRs}
   &&R=\frac{3}{4}e^{-f}\bigg(-2\tensor{\eta}{^{\mu\lambda}}\tensor{\p}{_{\mu}}\tensor{\p}{_\lambda}f + \tensor{\eta}{^{\mu\lambda}}\tensor{\p}{ _{\mu}}f\tensor{\p}{_\lambda}f\bigg)\\\label{confflatcurvSs}
   &&S=\frac{3}{4}e^{-f}\bigg(-2\tensor{\eta}{^{a b}}\tensor{\bar\p}{_{a}}\tensor{\bar\p}{_{b}}f+\eta^{a b}\tensor{\bar\p}{_{a}}f\tensor{\bar\p}{_{b}}f\bigg)
\end{eqnarray}
It is worth noting that these relations (\ref{confflatcurvtr}-\ref{confflatcurvSs}) show that although the underlying Minkowski tangent bundle is flat, the transformed spacetime has in general non-zero curvature. Thus, as is the case with a conformally flat Riemannian manifold, the anisotropic conformal flat tangent bundle is not flat itself. Nonetheless, each point has a neighborhood that can be mapped to a flat space by a conformal transformation, simply for example by the inverse transformation $e^{-f}$ of the metric. What distinguishes this case from the Riemannian one, however, is the non-zero $S-$curvature. This extra curvature, which, as shall be seen bellow, corresponds to extra gravity, could allow Finsler and Finsler-like structures to better describe the intricate nature of certain gravitational phenomena \cite{Kapsabelis2022, KOSTELECKY11, FOSTER15, hama2022dark, hama2021cosmological, hohmann2017geodesics, hohmann2019finsler}. Physically, in this model, the conformal factor represents the gravitational influence of dark matter. It can therefore be seen that, in this case, it is dark matter that curves the space-time even though the underlying base manifold is flat as is evident for example by relations (\ref{confflatcurvR}) and (\ref{confflatcurvS}). A particularly interesting special case of anisotropic conformal Minkowski spacetime arises if the conformal factor is solely direction dependent; namely if the conformal factor is of the form $f(y)$. Then it can clearly be seen that $R=0$ but in general $S\neq 0$, i.e. a flat horizontal space has nevertheless a non-flat tangent space since the vertical subspace is curved. This shows that the notion of "flatness" has been generalized on the tangent bundle since a seemingly flat space can still be anisotropic. 
\paragraph*{}Let us now consider the field equations (\ref{fieldeq}). Suppose the energy-momentum tensor field $\mathcal{T}$ is of the following form:
\begin{eqnarray}
&&\tensor{T}{_{\mu \nu}}=\begin{pmatrix}
\rho(x) & 0\\
0    & \tensor{g}{_{ij}}p(x)
\end{pmatrix}=
\begin{pmatrix}
\rho & 0\\
0    & e^{f}\tensor{I}{_3}p
\end{pmatrix}\\
&&\tensor{W}{_{a b}}=\begin{pmatrix}
\phi(x,y) & 0\\
0    & \tensor{g}{_{ij}}\psi(x,y)
\end{pmatrix}=
\begin{pmatrix}
 \phi & 0\\
0    & e^{f}\tensor{I}{_3}\psi
\end{pmatrix}
\end{eqnarray}
where $I_3=\diag(1,1,1)$ is the $3\times3$ identity matrix, $\rho(x)$, $p(x):M\rightarrow \mathbb{R}$ are the ordinary density and pressure functions of a thermodynamic fluid and $\phi(x,y)$, $\psi(x,y):TM\rightarrow \mathbb{R}$. These last two quantities; namely $\phi$ and $\psi$, could potentially be viewed as generalized thermodynamic variables on the tangent bundle\footnote{Further study of the thermodynamic properties of the vertical energy-momentum tensor is beyond the scope of this paper.}. Since the Ricci tensors given by (\ref{confflatcurvR}) and (\ref{confflatcurvS}) are not diagonal, we get the following non-trivial differential equations from the non-diagonal components of (\ref{fieldeq}):
\begin{eqnarray}\label{flatfieldh}
    2\tensor{f}{_{\mu\nu}}=\tensor{f}{_{\mu}}\cdot\tensor{f}{_{\nu}}\, , \, \forall \mu \neq \nu\\
    \label{flatfieldv}
    2\tensor{f}{_{ab}}=\tensor{f}{_{a}}\cdot\tensor{f}{_{b}}\, , \, \forall a \neq b
\end{eqnarray}
From the diagonal components we get the following set of 8 differential equations:
\begin{eqnarray}
    6\bigg(\tensor{f}{_{11}}+\tensor{f}{_{22}}+\tensor{f}{_{33}}\bigg)-3\bigg(( \tensor{f}{_{1}})^2+( \tensor{f}{_{2}})^2+( \tensor{f}{_{3}})^2\bigg)=10\rho-2\phi+6e^f(p+\psi)\label{flatfield1}\\
    6\bigg(\tensor{f}{_{00}}-\tensor{f}{_{11}}-\tensor{f}{_{22}}\bigg)+3\bigg(-( \tensor{f}{_{0}})^2+( \tensor{f}{_{1}})^2+( \tensor{f}{_{2}})^2\bigg) =2(\rho+\phi)-6e^f(p-\psi)\label{flatfield2}\\
    6\bigg(\tensor{f}{_{00}}-\tensor{f}{_{22}}-\tensor{f}{_{33}}\bigg)+3\bigg(-( \tensor{f}{_{0}})^2+( \tensor{f}{_{2}})^2+( \tensor{f}{_{3}})^2\bigg)=2(\rho+\phi)-6e^f(p-\psi)\label{flatfield3}
\end{eqnarray}
\begin{eqnarray}
    6\bigg(\tensor{f}{_{00}}-\tensor{f}{_{11}}-\tensor{f}{_{33}}\bigg)+3\bigg(-( \tensor{f}{_{0}})^2+( \tensor{f}{_{1}})^2+( \tensor{f}{_{3}})^2\bigg)\label{flatfield4}=2(\rho+\phi)-6e^f(p-\psi)\\
    6\bigg(\tensor{f}{_{55}}+\tensor{f}{_{66}}+\tensor{f}{_{77}}\bigg)-3\bigg(( \tensor{f}{_{5}})^2+( \tensor{f}{_{6}})^2+( \tensor{f}{_{7}})^2\bigg)\label{flatfield5}=10\phi-2\rho+6e^f(p+\psi)\\
    6\bigg(\tensor{f}{_{44}}-\tensor{f}{_{55}}-\tensor{f}{_{66}}\bigg)+3\bigg(-( \tensor{f}{_{4}})^2+( \tensor{f}{_{5}})^2+( \tensor{f}{_{6}})^2\bigg)\label{flatfield6}=2(\rho+\phi)+6e^f(p-\psi)\\
    6\bigg(\tensor{f}{_{44}}-\tensor{f}{_{66}}-\tensor{f}{_{77}}\bigg)+3\bigg(-( \tensor{f}{_{4}})^2+( \tensor{f}{_{6}})^2+( \tensor{f}{_{7}})^2\bigg)\label{flatfield7}=2(\rho+\phi)+6e^f(p-\psi)\\
    6\bigg(\tensor{f}{_{44}}-\tensor{f}{_{55}}-\tensor{f}{_{77}}\bigg)+3\bigg(-( \tensor{f}{_{4}})^2+( \tensor{f}{_{5}})^2+( \tensor{f}{_{7}})^2\bigg)\label{flatfield8}=2(\rho+\phi)+6e^f(p-\psi)
\end{eqnarray}
\paragraph*{}We shall now try to find a possible form for the conformal factor $f(x,y)$ using the field equations (\ref{flatfieldh}-\ref{flatfield8}). First, let $F:TM\longrightarrow\mathbb{R}_{>0}$ be some auxiliary function such that $\tensor{F}{_{\mu\nu}}=\tensor{F}{_{ab}}=0$ $\forall \mu\neq\nu$ and $a\neq b$. Such a function exists. For example:
\begin{equation}\label{polynomialsol}
    F(x,y)=\lambda_0(x^0)+\lambda_1(x^1)+\cdots+\lambda_7(x^7=y^3)
\end{equation}
with $\lambda_i:I\subseteq \mathbb{R}\rightarrow\mathbb{R}$, $\forall i=0,1,\cdots,7$ is such a function. Then $f(x,y)=-2\ln(F(x,y))$ satisfies equations (\ref{flatfieldh}) and (\ref{flatfieldv}). Next, by manipulating relations (\ref{flatfield2}-\ref{flatfield4}) and (\ref{flatfield6}-\ref{flatfield8}) we get the following relations:
\begin{eqnarray}\label{man1}
    2(\tensor{f}{_{11}}-\tensor{f}{_{22}})=(\tensor{f}{_{1}})^2-(\tensor{f}{_{2}})^2\\
    2(\tensor{f}{_{11}}-\tensor{f}{_{33}})=(\tensor{f}{_{1}})^2-(\tensor{f}{_{3}})^2\\
    2(\tensor{f}{_{22}}-\tensor{f}{_{33}})=(\tensor{f}{_{2}})^2-(\tensor{f}{_{3}})^2\\
    2(\tensor{f}{_{55}}-\tensor{f}{_{66}})=(\tensor{f}{_{5}})^2-(\tensor{f}{_{6}})^2\\
    2(\tensor{f}{_{55}}-\tensor{f}{_{77}})=(\tensor{f}{_{5}})^2-(\tensor{f}{_{7}})^2\\\label{man6}
    2(\tensor{f}{_{66}}-\tensor{f}{_{77}})=(\tensor{f}{_{6}})^2-(\tensor{f}{_{7}})^2
\end{eqnarray}
    Substituting $f=-2\ln{(F)}$ in relations (\ref{man1}-\ref{man6}) we get the following pieces of information concerning $F(x,y)$:
\begin{eqnarray}\label{p1}
    \tensor{F}{_{11}}=\tensor{F}{_{22}}=\tensor{F}{_{33}}\\\label{p2}
    \tensor{F}{_{55}}=\tensor{F}{_{66}}=\tensor{F}{_{77}}
\end{eqnarray}
Subsequently, using equations (\ref{flatfield1}), (\ref{flatfield5}),  (\ref{p1}) and (\ref{p2}) we arrive at the following equation:
\begin{equation}\label{geq}
    \frac{\tensor{F}{_{11}}-\tensor{F}{_{55}}}{F}=-\frac{1}{3}\big(\rho-\phi\big)
\end{equation}
Though it does not represent a general solution, equation (\ref{geq}) demonstrates the dependence of the conformal factor on the thermodynamic variables of the energy-momentum tensor. If, in addition, the auxiliary function $F(x,y)$ is of the form given in (\ref{polynomialsol}), then $\tensor{F}{_{11}}-\tensor{F}{_{55}}$ is a constant due to equations (\ref{p1}) and (\ref{p2}). Then:
\begin{equation} \label{conffla}
    \tensor{g}{_{\mu \nu}}=\chi (\rho-\phi)^2 \tensor{\eta}{_{\mu\nu}}
\end{equation}
where $\chi$ is a positive constant. In this special case (\ref{conffla}) it can clearly be seen that the conformal factor is connected with the distribution of energy and matter in the anisotropic conformal Minkowski spacetime.
\section{Anisotropic Conformal FLRW-Cosmology}\label{secFRW}
\paragraph*{}In this section we shall study an application of this geometric framework in cosmology. In particular, we shall use a FLRW metric structure for the underlying manifold $M$ and derive Friedmann-like equations of the horizontal subspace on the tangent bundle. 
\paragraph*{}Let 
\beqa
\tensor{\gamma}{_{\mu \nu}}= \begin{pmatrix}
-1 & 0 &0 &0\\
0 & \frac{a^2}{1-\kappa r^2} &0 &0\\
0 & 0 & (ra)^2 &0\\
0 & 0 & 0 & (ra\sin{\theta})^2\\
\end{pmatrix}
\eeqa
In this case we shall consider an integrable non-linear connection. The diagonal components of the Ricci tensor will then be:
\begin{eqnarray}\label{FRWRicci0}
    &&\tensor{R}{_{00}}=-3\frac{\Ddot{a}}{a}+\frac{1}{2}\tensor{\mathcal{L}}{_{00}}+\frac{1}{4}\tensor{M}{_{00}}\\\
    \label{FRWRicci1} &&\tensor{R}{_{11}}=\frac{a\Ddot{a}+2\dot{a}^2+2\kappa}{1-\kappa r^2}+\frac{1}{2}\tensor{\mathcal{L}}{_{11}}+\frac{1}{4}\tensor{M}{_{11}}\\
    \label{FRWRicci2} &&\tensor{R}{_{22}}=r^2(a\Ddot{a}+2\dot{a}^2+2\kappa)+\frac{1}{2}\tensor{\mathcal{L}}{_{22}}+\frac{1}{4}\tensor{M}{_{22}}\\
    \label{FRWRicci3} &&\tensor{R}{_{33}}=r^2\sin^2{\theta}(a\Ddot{a}+2\dot{a}^2+2\kappa)+\frac{1}{2}\tensor{\mathcal{L}}{_{33}}+\frac{1}{4}\tensor{M}{_{33}}
\end{eqnarray}
where:
\begin{eqnarray}
    &&\tensor{\mathcal{L}}{_{00}}=\frac{2\kappa a^2 r}{(1-\kappa r^2)^2}\tensor{\delta}{_{00}}f+\frac{1-\kappa r^2}{a^2}\tensor{\delta}{_{11}}f+\frac{1}{r^2a^2}\tensor{\delta}{_{22}}f +\frac{1}{r^2a^2\sin^2{\theta}}\tensor{\delta}{_{33}}f-\frac{3\Dot{a}}{a}\tensor{\delta}{_{0}}f+\frac{2-\kappa r^2}{ra^2}\tensor{\delta}{_{1}}f+\frac{\cot{\theta}}{r^2a^2}\tensor{\delta}{_{2}}f \\
    &&\tensor{M}{_{00}}=2(\tensor{\delta}{_{0}}f)^2 - (1-\kappa r^2)\bigg(\frac{\delta_{1}f}{a}\bigg)^2 - \bigg(\frac{\tensor{\delta}{_{2}}f}{ra}\bigg)^2 - \bigg(\frac{\tensor{\delta}{_{3}}f}{ra\sin{\theta}}\bigg)^2\\
    &&\tensor{\mathcal{L}}{_{11}}=\frac{a^2\tensor{\delta}{_{00}}f}{1-\kappa r^2}-\frac{\tensor{\delta}{_{22}}f}{(1-\kappa r^2)r^2}-\frac{\tensor{\delta}{_{33}}f}{r^2\sin^2{\theta}(1-\kappa r^2)} +\frac{4a\dot{a}}{1-\kappa r^2}\tensor{\delta}{_{0}}f-\frac{2}{r}\tensor{\delta}{_{1}}f-\frac{\cot{\theta}}{r^2(1-\kappa r^2)}\tensor{\delta}{_{2}}f \\
    &&\tensor{M}{_{11}}= -\frac{a^2(\tensor{\delta}{_{0}}f)^2}{1-\kappa r^2}+2(\tensor{\delta}{_{1}}f)^2+\frac{(\tensor{\delta}{_{2}}f)^2}{(1-\kappa r^2)r^2}+ \frac{(\tensor{\delta}{_{3}}f)^2}{r^2\sin^2{\theta}(1-\kappa r^2)}\\
    &&\tensor{\mathcal{L}}{_{22}}=r^2a^2\tensor{\delta}{_{00}}f-r^2(1-\kappa r^2)\tensor{\delta}{_{11}}f-\frac{\tensor{\delta}{_{33}}f}{\sin^2{\theta}} +4r^2a\Dot{a}\tensor{\delta}{_{0}}f-r(3-4\kappa r^2)\tensor{\delta}{_{1}}f -\cot{\theta}\tensor{\delta}{_{3}}f \\
    &&\tensor{M}{_{22}}= -r^2a^2(\tensor{\delta}{_{0}}f)^2+r^2(1-\kappa r^2) (\tensor{\delta}{_{1}}f)^2+2(\tensor{\delta}{_{2}}f)^2+ \frac{(\tensor{\delta}{_{3}}f)^2}{\sin^2{\theta}} \\
    &&\tensor{\mathcal{L}}{_{33}}=r^2a^2\sin^2{\theta}\tensor{\delta}{_{00}}f-r^2\sin^2{\theta}(1-\kappa r^2)\tensor{\delta}{_{11}}f -\sin^2{\theta}\tensor{\delta}{_{22}}f +4r^2a\Dot{a}\sin^2{\theta}\tensor{\delta}{_{0}}f -r(3-4\kappa r^2)\sin^2{\theta}\tensor{\delta}{_{1}}f-2\sin{\theta}\cos{\theta}\tensor{\delta}{_{2}}f\nn\\ \\
    &&\tensor{M}{_{33}}= -r^2a^2\sin^2{\theta}(\tensor{\delta}{_{0}}f)^2+r^2\sin^2{\theta}(1-\kappa r^2) (\tensor{\delta}{_{1}}f)^2 +\sin^2{\theta}(\tensor{\delta}{_{2}}f)^2 +2(\tensor{\delta}{_{3}}f)^2
\end{eqnarray}
It can be seen in equations (\ref{FRWRicci0}-\ref{FRWRicci3}) that the horizontal Ricci curvature fields of the tangent bundle are the form given in (\ref{Riccigentens}), i.e. they consist of the Ricci curvature $K$ of the Riemannian base manifold which appears perturbed by the two terms $\mathcal{L}$ and $M$ added to it. In particular, the Ricci curvature $K$ of the Riemannian base manifold is naturally independent of the conformal factor and hence free of the influence of dark gravity, while $\mathcal{L}$ is a pertubation of first order that is linear in terms of the conformal factor and $M$ is a second order pertubation which is non-linear in terms of the conformal factor. 
\paragraph*{}In general the horizontal Ricci tensor field $R$ is non-diagonal and non-symmetric, i.e. for $\mu \neq \nu$, $\tensor{R}{_{\mu \nu}}\neq 0$ and $\tensor{R}{_{\mu \nu}}\neq \tensor{R}{_{\nu \mu}}$. Due to this increased complexity in the geometric structure, we are going to limit this first order approach to the study of the diagonal terms (\ref{FRWRicci0}-\ref{FRWRicci3}). Although limited to these terms, important results could still be deduced from their study as a first order generalization of the classical theory of gravity on the tangent bundle.
\paragraph*{}The vertical Ricci curvature shall be:
\begin{eqnarray}
    &&\tensor{S}{_{00}}=\frac{1}{2}\bigg(\frac{1-\kappa r^2}{a^2}\tensor{\bar\p}{_{11}}f+ \frac{\tensor{\bar\p}{_{22}}f}{r^2a^2}+\frac{\tensor{\bar\p}{_{33}}f}{r^2 a^2 \sin^2{\theta}}\bigg)-\frac{1}{4}\bigg(\frac{1-\kappa r^2}{a^2}(\tensor{\bar\p}{_{1}}f)^2+ \frac{(\tensor{\bar\p}{_{2}}f)^2}{r^2a^2}+\frac{(\tensor{\bar\p}{_{3}}f)^3}{r^2 a^2 \sin^2{\theta}}\bigg)\\
    &&\tensor{S}{_{11}}=\frac{1}{2}\bigg(\frac{a^2\tensor{\bar\p}{_{00}}f}{1-\kappa r^2}-\frac{\tensor{\bar\p}{_{22}}f}{r^2(1-\kappa r^2)}-\frac{\tensor{\bar\p}{_{33}}f}{r^2\sin^2{\theta}(1-\kappa r^2)}\bigg)-\frac{1}{4}\bigg(\frac{(a\tensor{\bar\p}{_{0}}f)^2}{1-\kappa r^2}-\frac{(\tensor{\bar\p}{_{2}}f)^2}{r^2(1-\kappa r^2)}-\frac{(\tensor{\bar\p}{_{3}}f)^2}{r^2\sin^2{\theta}(1-\kappa r^2)}\bigg)\nn\\ \\
    &&\tensor{S}{_{22}}=\frac{1}{2}\bigg(r^2a^2\tensor{\bar\p}{_{00}}f-r^2(1-\kappa r^2)\tensor{\bar\p}{_{11}}f-\frac{\tensor{\bar\p}{_{33}}f}{\sin^2{\theta}}\bigg)-\frac{1}{4}\bigg((ra\tensor{\bar\p}{_{0}}f)^2-(1-\kappa r^2)(r\tensor{\bar\p}{_{1}}f)^2-\frac{(\tensor{\bar\p}{_{3}}f)^2}{\sin^2{\theta}}\bigg) \\
    &&\tensor{S}{_{33}}=\frac{1}{2}\bigg((ra\sin{\theta})^2\tensor{\bar\p}{_{00}}f-(r\sin{\theta})^2(1-\kappa r^2)\tensor{\bar\p}{_{11}}f-\sin^2{\theta}\tensor{\bar\p}{_{22}}f\bigg)
    \nn \\
    &&-\frac{1}{4}\bigg((ra\sin{\theta}\tensor{\bar\p}{_{0}}f)^2-(1-\kappa r^2)(r\sin{\theta}\tensor{\bar\p}{_{1}}f)^2-(\sin{\theta}\tensor{\bar\p}{_{2}}f)^2\bigg)\label{nondiagS}\\
    &&\tensor{S}{_{bc}}=\frac{1}{4}\bigg(2\tensor{\bar\p}{_{cb}}f-\tensor{\bar\p}{_{c}}f\bar\p_{b}f\bigg) ~,~ \forall b\neq c
\end{eqnarray}
As is the case with the horizontal Ricci curvature, the vertical Ricci is not diagonal but it is symmetric as is evident by relation (\ref{nondiagS}) if the conformal factor $f$ is at least $C^2$.
\paragraph*{}The scalar curvature will then be:
\begin{equation}
    \mathcal{R}=R+S=e^{-f}\bigg(K+\frac{1}{2}\mathcal{L}+\frac{1}{4}M+\frac{3}{4}\tilde{S}\bigg)
\end{equation}
where
\begin{eqnarray}
   &&R=e^{-f}\bigg(K+\frac{1}{2}\mathcal{L} +\frac{1}{4}M\bigg) \\
   &&S=\frac{3}{4}e^{-f}\tilde{S} \label{Cosm}\\
   &&K=6\bigg[ \frac{\ddot{a}}{a}+\bigg(\frac{\dot{a}}{a}\bigg)^2 +\frac{\kappa}{a^2} \bigg]\\
   &&\mathcal{L}=3\bigg[\tensor{\delta}{_{00}}f-\frac{1-\kappa r^2}{a^2}\tensor{\delta}{_{11}}f-\frac{\tensor{\delta}{_{22}}f}{(ra)^2}-\frac{\tensor{\delta}{_{33}}f}{(ra\sin{\theta})^2}+3\frac{\dot{a}}{a}\tensor{\delta}{_0}f -\frac{2-3\kappa r^2}{ra^2}\tensor{\delta}{_1}f -\frac{\cot{\theta}}{(ra)^2}\tensor{\delta}{_2}f\bigg] \\
   &&M=-3\bigg[(\tensor{\delta}{ _{0}}f)^2-\frac{1-\kappa r^2}{a^2}(\tensor{\delta}{ _{1}}f)^2-\frac{(\tensor{\delta}{ _{2}}f)^2}{(ra)^2}-\frac{(\tensor{\delta}{ _{3}}f)^2}{(ra\sin{\theta})^2}\bigg] \\
   &&\tilde{S}=-2\bigg[\tensor{\bar\p}{ _{00}}f-\frac{1-\kappa r^2}{a^2}\tensor{\bar\p}{ _{11}}f-\frac{\tensor{\bar\p}{ _{22}}f}{(ra)^2}-\frac{\tensor{\bar\p}{ _{33}}f}{(ra\sin{\theta})^2}\bigg]-(\tensor{\bar\p}{ _{0}}f)^2+\frac{1-\kappa r^2}{a^2}(\tensor{\bar\p}{ _{1}}f)^2+\frac{(\tensor{\bar\p}{ _{2}}f)^2}{(ra)^2}+\frac{(\tensor{\bar\p}{ _{3}}f)^2}{(ra\sin{\theta})^2} \nn \\ 
\end{eqnarray}

\subsection{Extended anisotropic conformal Friedmann-like equations}
\paragraph*{}Let us now consider the field equations (\ref{fieldzero}). Suppose, as in section \ref{secflat}, that the energy-momentum tensor field $\mathcal{T}$ is of the following form:
\begin{eqnarray}
&&\tensor{T}{_{\mu \nu}}=\begin{pmatrix}
\rho(x) & 0\\
0    & \tensor{g}{_{ij}}p(x)
\end{pmatrix}\\
&&\tensor{W}{_{a b}}=\begin{pmatrix}
\phi(x,y) & 0\\
0    & \tensor{g}{_{ij}}\psi(x,y)
\end{pmatrix}
\end{eqnarray}
where $\rho(x)$, $p(x):M\rightarrow \mathbb{R}$ are the ordinary density and pressure functions of a thermodynamic fluid and $\phi(x,y)$, $\psi(x,y):TM\rightarrow \mathbb{R}$ could potentially be viewed as generalized thermodynamic variables on the tangent bundle. Then, the horizontal Friedmann-like equations shall be of the following form:
\begin{eqnarray}
    &&\bigg(\frac{\dot{a}}{a}\bigg)^2+\frac{\kappa}{a^2}+\frac{1}{8}\tilde{S}+\frac{1}{12}
    X+\frac{1}{24}\Phi=\frac{\rho}{3}\\ \label{PreHorizontalFriedmann}
    &&\frac{\ddot{a}}{a}+\frac{1}{8}\tilde{S}-\frac{1}{12}X_i-\frac{1}{24}\Phi_i=-\frac{\rho+3e^fp}{6}
\end{eqnarray}
where $i=1,2,3$ and 
\begin{eqnarray}
    &&X=\frac{3+4\kappa a^2 r-6\kappa r^2+3\kappa^2 r^4}{(1-\kappa r^2)^2}\tensor{\delta}{_{00}}f-\frac{1-\kappa r^2}{a^2}\tensor{\delta}{_{11}}f -\frac{\tensor{\delta}{_{22}}f}{(ra)^2}-\frac{\tensor{\delta}{_{33}}f}{(ra\sin{\theta})^2}+3\frac{\dot{a}}{a}\tensor{\delta}{_{0}}f-\frac{2-7\kappa r^2}{ra^2}\tensor{\delta}{_{1}}f-\frac{\cot{\theta}}{(ra)^2}\tensor{\delta}{_{2}}f \label{chi}\nn \\  \\
    &&\Phi=(\tensor{\delta}{_{0}}f)^2+\frac{1-\kappa r^2}{a^2}(\tensor{\delta}{_{1}}f)^2+\frac{(\tensor{\delta}{_{2}}f)^2}{(ra)^2}+\frac{(\tensor{\delta}{_{3}}f)^2}{(ra\sin{\theta})^2} \nn \\ \label{phi}\\
    &&X_1=\frac{2\kappa a^2 r}{(1-\kappa r^2)^2}\tensor{\delta}{_{00}}f+4\frac{1-\kappa r^2}{a^2}\tensor{\delta}{_{11}}f +\frac{\tensor{\delta}{_{22}}f}{(ra)^2}+\frac{\tensor{\delta}{_{33}}f}{(ra\sin{\theta})^2}+\frac{2-4\kappa r^2}{ra^2}\tensor{\delta}{_{1}}f+\frac{\cot{\theta}}{(ra)^2}\tensor{\delta}{_{2}}f \\
    &&\Phi_1=2(\tensor{\delta}{_{0}}f)^2+2\frac{1-\kappa r^2}{a^2}(\tensor{\delta}{_{1}}f)^2-\frac{(\tensor{\delta}{_{2}}f)^2}{(ra)^2}-\frac{(\tensor{\delta}{_{3}}f)^2}{(ra\sin{\theta})^2}\\
    &&X_2=\frac{2\kappa a^2 r}{(1-\kappa r^2)^2}\tensor{\delta}{_{00}}f+\frac{1-\kappa r^2}{a^2}\tensor{\delta}{_{11}}f+4\frac{\tensor{\delta}{_{22}}f}{(ra)^2} +\frac{\tensor{\delta}{_{33}}f}{(ra\sin{\theta})^2}-\frac{1-2\kappa r^2}{ra^2}\tensor{\delta}{_{1}}f+\frac{\cot{\theta}}{(ra)^2}\tensor{\delta}{_{2}}f\\
    &&\Phi_2=2(\tensor{\delta}{_{0}}f)^2-\frac{1-\kappa r^2}{a^2}(\tensor{\delta}{_{1}}f)^2+2\frac{(\tensor{\delta}{_{2}}f)^2}{(ra)^2}-\frac{(\tensor{\delta}{_{3}}f)^2}{(ra\sin{\theta})^2} \\
    &&X_3=\frac{2\kappa a^2 r}{(1-\kappa r^2)^2}\tensor{\delta}{_{00}}f+\frac{1-\kappa r^2}{a^2}\tensor{\delta}{_{11}}f+\frac{\tensor{\delta}{_{22}}f}{(ra)^2}+4\frac{\tensor{\delta}{_{33}}f}{(ra\sin{\theta})^2}-\frac{1-2\kappa r^2}{ra^2}\tensor{\delta}{_{1}}f-2\frac{\cot{\theta}}{(ra)^2}\tensor{\delta}{_{2}}f\\
    &&\Phi_3=2(\tensor{\delta}{_{0}}f)^2-\frac{1-\kappa r^2}{a^2}(\tensor{\delta}{_{1}}f)^2-\frac{(\tensor{\delta}{_{2}}f)^2}{(ra)^2}+2\frac{(\tensor{\delta}{_{3}}f)^2}{(ra\sin{\theta})^2}
\end{eqnarray}
By virtue of the three equations (\ref{PreHorizontalFriedmann}) we get the following pair of generalized anisotropic conformal Friedmann-like equations for the horizontal subspace on the tangent bundle:
\begin{eqnarray}\label{HorizontalFriedmann1}
    &&\bigg(\frac{\dot{a}}{a}\bigg)^2+\frac{\kappa}{a^2}+\frac{1}{8}\tilde{S}+\frac{1}{12}
    X+\frac{1}{24}\Phi=\frac{\rho}{3}\\ \label{HorizontalFriedmann2}
    &&\frac{\ddot{a}}{a}+\frac{1}{8}\tilde{S}-\frac{1}{6}\Psi-\frac{1}{12}\mathfrak{D}=-\frac{\rho+3e^fp}{6}
\end{eqnarray}
where $X$ and $\Phi$ are given in relations (\ref{chi},\ref{phi}) and 
\begin{eqnarray}
    &&\Psi=\frac{\kappa a^2 r}{(1-\kappa r^2)^2}\tensor{\delta}{_{00}}f+\frac{1-\kappa r^2}{a^2}\tensor{\delta}{_{11}}f +\frac{\tensor{\delta}{_{22}}f}{(ra)^2}+\frac{\tensor{\delta}{_{33}}f}{(ra\sin{\theta})^2}\\
    &&\mathfrak{D}=(\tensor{\delta}{_{0}}f)^2
\end{eqnarray}
In view of relations (\ref{HorizontalFriedmann1},\ref{HorizontalFriedmann2}), it is worth noting that the generalized anisotropic conformal Friedmann-like equations for the horizontal subspace on the tangent bundle include extra terms denoted by $X$, $\Psi$, $\Phi$, $\mathfrak{D}$ which introduce a higher order structure derived by the gravitational influence of dark matter and dark energy, which enrich the cosmological study of the evolution of the universe with further information. It is also clear that if these terms are equal to zero then equations (\ref{HorizontalFriedmann1},\ref{HorizontalFriedmann2}) reduce to the ordinary Friedmann equations of general relativity. In particular, it can be seen that the scalar Ricci curvature $S$ of the vertical subspace and especially $\tilde{S}$, could be related to a dynamical anisotropic cosmological "constant" as is shown in \cite{Triantafyllopoulos_2018}, which emerges from the additional degrees of freedom of the anisotropic conformal geometric structure instead of being added ad hoc as in the classical case. Therefore, equations (\ref{HorizontalFriedmann1},\ref{HorizontalFriedmann2}) reduce to the Friedmann equations of general relativity with dynamical cosmological parameter equal to $\tilde{S}=-\frac{8}{3}\Lambda$, where in this case $\Lambda$ denotes the varying cosmological constant \cite{Papagiannopoulos2020}. By means of relation (\ref{Cosm}) the cosmological parameter is related to the scalar vertical curvature $S$ in precisely the same way as in \cite{Triantafyllopoulos_2018}. With respect to the classical case, the presence of a varying cosmological constant in the form of $\tilde{S}$ indicates a different dynamical evolution of the universe which could be compared to the $\Lambda-$CDM cosmological model \cite{Papagiannopoulos2020} in a further study, possibly viewed through the lens of a mimetic dark gravity model. In general, if $f=0$, i.e. in the absence of dark gravity, then (\ref{HorizontalFriedmann1},\ref{HorizontalFriedmann2}) reduce exactly to the classical Friedmann equations (without cosmological constant), as well as to the geometric frame as described in \cite{Triantafyllopoulos_2018} for a flat vertical subspace.

 A special case of conformal factors that are of the form $f(x,y^0)$ is of noteworthy interest since, this family of conformal transformations leave the vertical subspace isotropic in the sense presented in \cite{carroll2004spacetime}, i.e. the vertical $S$-curvature tensor is diagonal and $\tensor{S}{_{ij}}=\frac{1}{4}\tensor{\gamma}{_{ij}}(2\tensor{\bar\p}{_{00}}f-(\tensor{\bar\p}{_{0}} f)^2)$. In particular $\tensor{S}{_{00}}=0$. The vertical field equation (\ref{fieldeq}) for $(a=b=0)$ then yields:
\begin{equation}\label{spfrw}
    \rho-5\phi=3e^f(p+\psi)\xrightarrow{\mathrm{if}~p\neq-\psi}f(x,y^0)=\ln{\bigg(\frac{\rho-5\phi}{3(p+\psi)}\bigg)}
\end{equation}
As is the case in (\ref{conffla}) it can be seen that the conformal factor of relation (\ref{spfrw}) is connected with the distribution of energy and matter in the anisotropic conformal spacetime. This is another indication that the conformal factor may be related to the thermodynamic properties of the spacetime.\\
\begin{center}{\begin{tikzcd}
	\text{DM} \\
	& {f(x,y)} & S & \Lambda \\
	\text{DE} \\
	& {\text{Graph 1}}
	\arrow["{?}"{description}, tail reversed, from=1-1, to=3-1]
	\arrow[from=1-1, to=2-2]
	\arrow[from=3-1, to=2-2]
	\arrow[from=2-2, to=2-3]
	\arrow[from=2-3, to=2-4]
	\arrow[curve={height=12pt}, tail reversed, from=3-1, to=2-4]
\end{tikzcd}}
\end{center}
\paragraph*{}Graph 1 summarizes the relations between the physical and geometrical concepts that arise from this section. In particular, the internal properties (possibly of thermodynamical nature) of DM and DE are mathematically expressed through the conformal factor $f$ which in turn induces a vertical $S$-curvature. This curvature produces then a varying dynamical cosmological "constant" $\Lambda$ which is related to DE. A relation between DE and DM could possibly be studied as discussed in the next section \ref{conclusion}. \\
\paragraph*{} In light of the generalized anisotropic conformal Friedmann-like equations for the horizontal subspace on the tangent bundle (\ref{HorizontalFriedmann1}-\ref{HorizontalFriedmann2}), we can obtain the following pair of equations for the Hubble parameter $H(t):=\dot{a}/a$, as follows
\begin{eqnarray}\label{HorizontalFriedmannHubble1}
    &&3H^2+\frac{3\kappa}{a^2}=\rho+\rho_{DE}\\ \label{HorizontalFriedmannHubble2}
    &&2\dot{H}+3H^2+\frac{\kappa}{a^2}=-(e^fp+p_{DE})
\end{eqnarray}
where
\begin{eqnarray}\label{DErho}
    &&\rho_{DE}:=-\frac{3}{8}\tilde{S}-\frac{1}{4}
    X-\frac{1}{8}\Phi\\\label{DEp}
    &&p_{DE}:=\frac{3}{8}\tilde{S}+\frac{1}{12}
    X+\frac{1}{24}\Phi-\frac{\Psi}{3}-\frac{1}{6}\mathfrak{D}
\end{eqnarray}
could be interpreted as the density and pressure of DE, respectively. Let us now consider the following special case of a simplified linear conformal factor, i.e. let $f$ be of the form:
\begin{equation}
    f(x,y)=\beta t +\mu y^0+\nu y^1.
\end{equation}
Additionally, let us focus on a spatially flat universe and assume dust matter; namely let $\kappa=0$ and $p=0$. For the simplicity, let the coefficients of the non-linear vanish identically. In this case, we obtain $X=-3\beta H $, $\Phi=\beta^2$, $\tilde{S}=-\mu^2+\frac{\nu^2}{a^2}$, $\Psi=0$ and  $\mathfrak{D}=\beta^2$,    
    where $H(t)= \dot{a}/a$ is the Hubble function. Then from the Friedmann-like equations (\ref{HorizontalFriedmannHubble1}-\ref{HorizontalFriedmannHubble2}) we get:
\begin{eqnarray}\label{HorizontalFriedmann1bb}
3H^2=\rho  +\rho_{DE}\\
\label{HorizontalFriedmann2bb}
2\dot{H}+3H^2=-p_{DE}
\end{eqnarray}
where 
\begin{eqnarray}
&&\rho_{DE}= \frac{1}{8}(3\mu^2-\beta^2)-\frac{3}{8}\frac{\nu^2}{a^2}+\frac{3}{4}\beta H
\label{rhoDE}\\
&&p_{DE}= -\frac{1}{8}(3\mu^2+\beta^2)+\frac{3}{8}\frac{\nu^2}{a^2}-\frac{1}{4}\beta H.
\label{pDE}
\end{eqnarray}
Hence, as described above, the richer structure of Finsler geometry produces an effective dark energy sector of geometric origin. The first term in (\ref{rhoDE}) is constant and accounts for the usual cosmological constant, the second term is an effective spatial curvature term that will have a negligible role at late-time universe, and the last term is a novel friction term. Additionally, since $p=0$ the evolution of $\rho$ reads simply $\rho=\rho(0) a^{-3}$, while we can define the effective dark-energy equation-of-state parameter as $w_{DE}\equiv p_{DE}/\rho_{DE}$.
Finally, we introduce the density parameters  $\Omega_{m}\equiv \frac{\rho}{3H^2}$ and $\Omega_{DE}\equiv \frac{\rho_{DE}}{3H^2}$, while we use the redshift $z$ as the independent variable, defined through $1+z=\frac{a_0}{a}$ (and setting the present scale factor as $a_0=1$).

\begin{figure}[!h]
\centering\vspace{-2.3cm}\hspace{-1.5cm}
\includegraphics[width=9cm,height=10.cm]{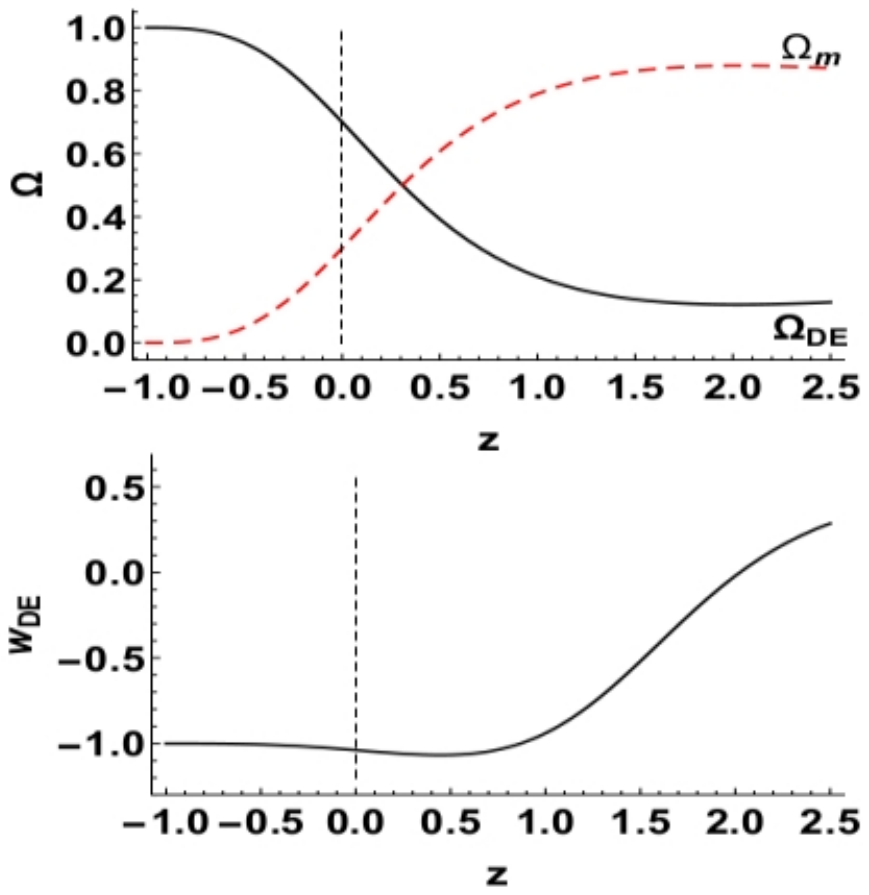} \vspace{-2.5cm}                     \caption{\it{Upper graph: The evolution of the effective dark energy density 
parameter $\Omega_{DE}$  and of the matter density parameter 
$\Omega_{m}$,   as a function of the redshift $z$, for $\mu=1$, $\beta=0.1$ and $\nu=0.1$. Lower graph: 
The evolution of the corresponding dark-energy 
equation-of-state parameter
$w_{DE}$.  We have imposed $ \Omega_{m0}\approx0.3$ at 
present time. 
}}
\label{Omegas}
\end{figure}

We evolve equations (\ref{HorizontalFriedmann1bb}),(\ref{HorizontalFriedmann2bb}) numerically and in Fig. \ref{Omegas} we depict the evolution of the effective dark energy density 
parameter $\Omega_{DE}$  and of the matter density parameter 
$\Omega_{m}$, as well as the evolution of the effective dark-energy 
equation-of-state.

\begin{figure}[ht]
  \centering\vspace{-0.5cm}\hspace{-1.5cm}
\includegraphics[width=9cm,height=8.cm]{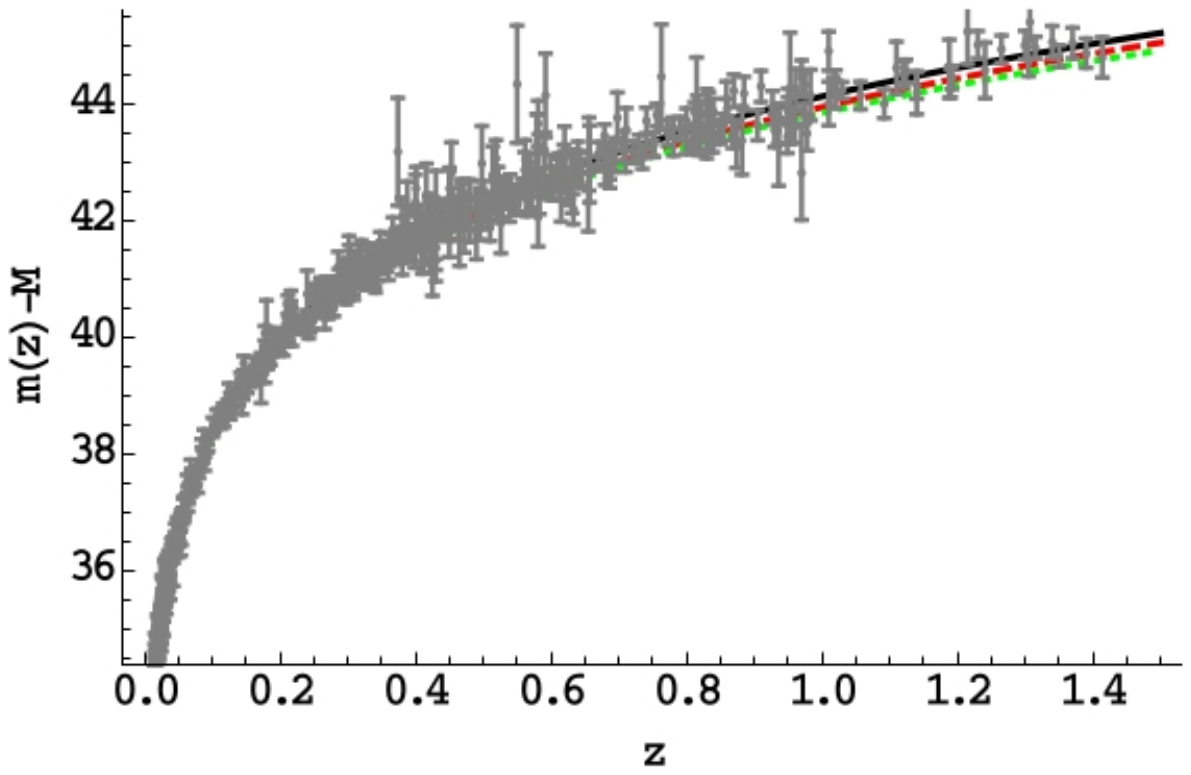}\vspace{-1.3cm}
\caption{
{\it{The theoretically predicted apparent minus absolute magnitude     for 
for $\mu=1$, $\beta=0.1$ and $\nu=0.1$ (red-dashed) and 
for $\mu=1$, $\beta=0.2$ and $\nu=0.3$(green-dotted). The observational points correspond to the  $580$ SN Ia data 
points from
\cite{SupernovaCosmologyProject:2011ycw}, and for completeness and comparison  we 
depict the prediction of  $\Lambda$CDM cosmology with the black-solid curve.
}} }
\label{Fig2}
\end{figure}

Finally, we perform a confrontation of the obtained $H(z)$ behavior with  Supernovae type Ia (SN Ia) data. In particular, one measures  the apparent 
magnitude $m(z)$ which is related to the luminosity distance as
$ 
  m(z) - M = 5 
\log\left[\frac{d_L(z)_{\text{obs}}}{Mpc}\right]  + 25,
$ 
with $M$ and $L$ the absolute magnitude and luminosity.
Moreover, the predicted   
luminosity distance $d_{L}(z)_\text{th}$ is given as
$
d_{L}\left(z\right)_\text{th}\equiv\left(1+z\right)
\int^{z}_{0}\frac{dz'}{H\left(z'\right)}$.
In Fig.   \ref{Fig2} we present the theoretically predicted apparent minus absolute 
magnitude    as well as the prediction of 
$\Lambda$CDM cosmology, on top of the   $580$ SN Ia observational data points from 
\cite{SupernovaCosmologyProject:2011ycw}. As we observe, the agreement   is excellent.  

\section{Concluding Remarks and some Future Prospects}\label{conclusion}
Motivated by the apparent need for a mathematical and in particular geometrical framework for a theory of gravity that includes the significant contributions of dark matter and dark energy, whose study currently constitutes the greatest problem of modern cosmology \cite{Bouali2023} we developed a theoretical model based on an anisotropic conformal spacetime on the tangent bundle that allows for extra degrees of freedom due to the higher-dimensionality of the underlying geometry which intrisically incorporates the concept of an anisotropic direction dependent dark gravity in the metric structure. This higher order internal geometric structure is interpreted as the contributions of dark matter and dark energy. In particular, in this framework we examined two cases of significant interest; namely the conformal anisotropic Minkowski spacetime and the conformal anisotropic FLRW-cosmology. 
\paragraph*{}A first application of this geometric framework is given by using a Minkowski metric structure for the underlying manifold. The interest of this case lies in its potential cosmological application for the study of a post-inflation universe which evolves towards flatness, because this metrical model might be connected to an anisotropic generalization of a de-Sitter metric spacetime in which a conformal structure for the spacelike metric, (e.g. in a Friedmann metric space) is caused by the scale factor. The study of the anisotropic conformal Minkowski space, especially rel. (\ref{confflatcurvR}, \ref{confflatcurvS}), reveals that even though the underlying base manifold is flat, dark matter, represented by the conformal factor, curves the space-time. A further study of special types of conformal factors that constitute solutions to the field equations for this spacetime, reveal a first indication in eq. (\ref{conffla}) that the conformal factor, which we nevertheless consider given a priori (e.g. determined by observational or experimental data), is potentially connected to the thermodynamic variables of energy and matter. A future dynamical analysis of this model, similar to that studied in \cite{Kapsabelis2022, Papagiannopoulos2020}, could provide critical points which are vital regions of the evolution of the universe. Relating these results with current observational data could lead to a more complete understanding of this anisotropic geometric framework of gravity and cosmology, as well as of the contribution of dark matter and dark energy to the evolution of the universe. In particular, a deeper study of the conformal factor could be performed by imposing certain extra physical conditions on this model which are consistent with observational data. For instance, one could assume that the horizontal space tends towards flatness for large time, i.e. that the R-curvature tensor given in eq. (\ref{confflatcurvtr}) tends towards zero as the time parameter tends to infinity. Furthermore, assuming that the anisotropy of the universe reduces for large time, i.e. that the universe tends towards isotropy, one could argue that the vertical S-curvature given in eq. (\ref{confflatcurvts}) should diminish in the limit as time tends to infinity. Due to the afore-mentioned connection of the conformal factor to the thermodynamic structure of energy and matter, the thermodynamic implications of such a future study could potentially be linked to the notion of a cosmological entropy \cite{Nojiri2019, Saridakis_2020}.
\paragraph*{}A second application studied in this work, is an anisotropic conformal FLRW space which is, naturally, of significant cosmological interest. According to the the generalized anisotropic conformal Friedmann-like equations for the horizontal subspace on the tangent bundle that we derived, rel. (\ref{HorizontalFriedmann1},\ref{HorizontalFriedmann2}), we find that the classical Riemannian structure appears perturbed by the inclusion of extra terms which arise naturally by the geometry of the tangent bundle, are linked to the higher order structure of this framework and are interpreted as the additional gravitational influence of dark matter and dark energy in the cosmological evolution of the universe. We find, in particular, that the classical Friedmann equations of general relativity, as well as the generalized Friedmann equations in \cite{Triantafyllopoulos_2018} with dynamical cosmological parameter can be recovered if we interpret the vertical scalar curvature $S$ as the varying cosmological constant in much the same way as in \cite{Triantafyllopoulos_2018}. This cosmological parameter is produced internally through the geometry of the tangent bundle instead of being added ad hoc as in the classical case and could potentially be quantitatively studied for a given conformal factor.One such first approach at a more concrete example of a simple conformal factor is provided and connected with observational constraints on both dark energy and dark matter. We find that this special case is very consistent with observational results as well as with $\Lambda$-CDM, which could suggest that this model is promising and that further work on combining recent observational data and constraints, as in \cite{HARKO20221}, with our theoretical model could potentially yield ever more accurate conformal factors to fit the observational results. In addition, a future study of the bounce conditions applied to this model could prove fruitful, as they could endow the conformal factor, and hence the contribution of dark matter and dark energy, with essential cosmological information related to the dynamic anisotropic evolution of the universe. For this purpose, a careful investigation of the equation of state of the anisotropic generalized thermodynamic variables of the cosmological fluid, and of the energy conditions that may ensue from the generalized anisotropic conformal Friedmann-like equations (\ref{HorizontalFriedmann1}, \ref{HorizontalFriedmann2}) might prove invaluable for a deeper understanding of the connection between dark gravity, cosmology and anisotropy.
\paragraph*{}In conclusion, this geometric framework of conformal gravity on the tangent bundle that incorporates the gravitational influence of dark matter and dark energy could allow for both a qualitative and a quantitative analysis of the cosmological aspects of the evolution of the universe, for instance in a future study that includes an application of this model using observational data. In particular, a quantitative study of the deflection angle using this model may provide for a correction due to dark gravity of the already known results given for an anisotropic Finsler-Randers model in \cite{Kapsabelis2022}. Moreover, potential links between dark energy and dark matter on galactic scales could be studied, since the behavior of the dark matter cosmological fluid on a large scale could reveal a relation with dark energy \cite{Anagnostopoulos_2019, Arbey, Arbey_2021}. Connected to such a future endeavour could possibly be the model of the Chaplygin gas \cite{chaplygin} whose study as a dark cosmological fluid instead of the perfect fluid model, in conjunction with the present geometric model could potentially yield interesting results. Finally, an especially interesting future prospect of this mathematical framework would be the application of the present work in the development of a galactic model combining the already existing dark matter halo theory \cite{Benetti_2023} and related observational data in order to study structure formations due to anisotropy. 
  \\
\section*{Acknowledgments}
First, we wish to thank the unknown referee(s) for their indispensable comments and suggestions that helped improve the present article. We would like to thank Dr. S. Konitopoulos for his insightful comments and fruitful discussions. We would also like to especially thank Dr. E.N. Saridakis for his invaluable help and astute comments. Finally, we thank Dr. F. K. Anagnostopoulos for our interesting discussions.

\appendix

\section{Appendix section}\label{app}
\paragraph*{}Let 
\beqa
\tensor{\gamma}{_{\mu \nu}}= \begin{pmatrix}
-1 & 0 &0 &0\\
0 & \frac{a^2}{1-\kappa r^2} &0 &0\\
0 & 0 & (ra)^2 &0\\
0 & 0 & 0 & (ra\sin{\theta})^2\\
\end{pmatrix}
\eeqa
Let the non-linear connection be of Cartan-type; namely we take $\tensor{N}{^a_\kappa}=\tensor{\gamma}{^a_{b\kappa }}y^b$. Then the components of the non linear connection are as follows:
\begin{eqnarray}
    &&\tensor{N}{^0_0}=0 \\
    &&\tensor{N}{^0_1}=\frac{a\dot{a}}{1-\kappa r^2}y^1 \\
    &&\tensor{N}{^0_2}=a\dot{a}r^2y^2\\
    &&\tensor{N}{^0_3}=a\dot{a}r^2\sin^2{\theta}y^3\\
    &&\tensor{N}{^1_0}=\frac{\dot{a}}{a}y^1 \\
    &&\tensor{N}{^1_1}=\frac{\dot{a}}{a}y^0+\frac{\kappa r}{1-\kappa r^2}y^1\\
    &&\tensor{N}{^1_2}=-r(1-\kappa r^2)y^2 \\
    &&\tensor{N}{^1_3}=-r(1-\kappa r^2)y^2\sin^2{\theta}y^3 \\
    &&\tensor{N}{^2_0}=\frac{\dot{a}}{a}y^2 \\
    &&\tensor{N}{^2_1}=\frac{1}{r}y^2 \\
    &&\tensor{N}{^2_2}=\frac{\dot{a}}{a}y^0+\frac{1}{r}y^1 \\
    &&\tensor{N}{^2_3}=-\sin{\theta}\cos{\theta}y^3 \\
    &&\tensor{N}{^3_0}=\frac{\dot{a}}{a}y^3 \\
    &&\tensor{N}{^3_1}=\frac{1}{r}y^3 \\
    &&\tensor{N}{^3_2}=\cot{\theta}y^3 \\
    &&\tensor{N}{^3_3}=\frac{\dot{a}}{a}y^0+\frac{1}{r}y^1+\cot{\theta}y^2
\end{eqnarray}
Indeed we can clearly see that such a non-linear connection is not integrable.

\bibliographystyle{unsrtnat}
\bibliography{apssamp.bib} 

\end{document}